\documentclass[aps,prd,amsmath,amssymb,showpacs]{revtex4}
\usepackage{graphicx}
\usepackage{dcolumn}
\usepackage{morefloats}
\usepackage{color}
\usepackage{slashed}
\usepackage{bbm}
\begin{document}
\title{Comprehensive study of light axial vector mesons with the presence of triangle singularity}

\author{Meng-Chuan Du$^{1,2}$\footnote{{\it E-mail address:} dumc@ihep.ac.cn},
and Qiang Zhao$^{1,2}$\footnote{{\it E-mail address:} zhaoq@ihep.ac.cn}
}
\affiliation{$^1$ Institute of High Energy Physics and Theoretical Physics Center
for Science Facilities,\\
         Chinese Academy of Sciences, Beijing 100049, P.R. China}

\affiliation{$^2$ University of Chinese Academy of
Sciences, Beijing 100049, P.R. China}

\begin{abstract}
We present a systematic study of the productions and decays of light axial vector mesons with $J^{PC}=1^{+\pm}$ in charmonium decays. In the quark model scenario, the two axial vector nonets are connected with each other by the Gell-Mann-Okubo mass relation through the mixing between two $K_1$ states (i.e. $K_1(1270)$ and $K_1(1400)$) with configurations of $^3P_1$ and $^1P_1$. The mixing angles between $f_1$ and $f_1'$ (i.e. $f_1(1285)$ and $f_1(1420)$), and between $h_1$ and $h_1'$ (i.e. $h_1(1170)$ and $h_1(1415)$) can be reliably constrained. We then introduce the intermediate $K^*\bar{K}+c.c.$ meson loop transitions in the description of the productions and decays of these axial vector mesons. The presence of the nearby $S$-wave $K^*\bar{K}+c.c.$ to which these axial vector mesons have strong coupling strengths, turns out to be crucial for understanding many puzzling questions related to their productions and decays. This is because that the $S$-wave $K^*\bar{K}+c.c.$ rescatterings by the kaon exchange satisfy the triangle singularity (TS) condition in some of these cases and the TS mechanism can introduce special interference effects in the exclusive decays of these light axial vector mesons.

\end{abstract}

\maketitle

\section{Introduction}

The constituent quark model has proved to be successful on the description of many qualitative features of low-lying hadrons. However, it also shows that it is challenging for us to provide a quantitative prescription for the hadron spectroscopy due to the complexity of non-perturbative QCD. A long-standing puzzling case is the properties of the lowest-lying positive parity vectors, namely, the axial vector mesons. With many unanswered mysterious questions on their productions and decays it seems that a consistent picture for the axial vector nonets as the first orbital excitation states in the light meson spectroscopy is still far from broadly accepted.

In the Particle Data Group (PDG)~\cite{Patrignani:2016xqp,Zyla:2020zbs} the following nonstrange light axial vector states are identified in experiment, namely, $a_1(1260)$, $f_1(1285)$, and $f_1(1420)$ as multiplets of $J^{PC}=1^{++}$, and $b_1(1235)$, $h_1(1170)$ and $h_1(1415)$~\footnote{The state $h_1(1415)$ was labelled as $h_1(1380)$ in the PDG before the 2020 Edition~\cite{Zyla:2020zbs}. The change is due to the new data for $J/\psi\to h_1'\eta'\to K\bar{K}\pi\eta'$ from BESIII, where the resonance parameters have been updated~\cite{Ablikim:2018ctf}. } as multiplets of $J^{PC}=1^{+-}$. Two strange multiplets, $K_1(1270)$ and $K_1(1400)$, are identified as their strange partners, respectively, and then two quark-model axial vector nonets can be constructed. Since $K_1(1270)$ and $K_1(1400)$ do not have a fixed $C$-parity their mixing has impact on the nonstrange axial vector meson masses through the Gell-Mann-Okubo relation, and it has been the focus of many studies in the literature. While the mixing angle between $K_1(1270)$ and $K_1(1400)$ has been well established during the past decade (see a brief review by Refs.~\cite{Cheng:2011pb,Cheng:2013cwa} and references therein), more puzzling but interesting issues were raised about the $f_1(1285)$ and $f_1(1420)$ and they have been proposed to be exotic states based on their couplings to the $K^*\bar{K}+c.c.$ open channel.

In fact, the role played by the $S$-wave $K^*\bar{K}+c.c.$ open threshold has been recognised in almost all the non-$q\bar{q}$ interpretations for the nonstrange axial vector mesons in the literature. For instance, it was proposed in Ref.~\cite{Longacre:1990uc} that $f_1(1420)$ could be a $K^*\bar{K}+c.c.$ molecule.
In Ref.~\cite{Roca:2005nm} it was found that the $S$-wave interactions between the vector and pseudoscalar mesons would lead to pole structures in the second Riemann sheet. Within this scenario axial vector states $f_1(1285)$, $a_1(1420)$, $h_1(1170)$, $h_1(1385)$ and $b_1(1235)$ were interpreted as dynamically generated states. In particular, $f_1(1285)$ was proposed to be a $K^*\bar{K}+c.c.$ molecule, while the authors found that $f_1(1420)$ could not be accommodated by their framework. In Ref.~\cite{Liang:2019vhf} an analysis of $J/\psi\to b_1(1235)\pi$, $h_1(1170)\eta^{(\prime)}$, and $h_1(1380)\eta^{(\prime)}$ was presented by treating the $C=-1$ axial vector mesons as dynamically generated states by the pseudoscalar and vector $S$-wave couplings. 

Although the $S$-wave $K^*\bar{K}+c.c.$ open threshold appears to be crucial for all the present non-$q\bar{q}$ interpretations, model-dependent assumptions in different phenomenologies make it difficult to distinguish those scenarios beyond the simple quark model $q\bar{q}$ categorizations. In this work we show that the $S$-wave $K^*\bar{K}+c.c.$ interactions actually cut in the problem via its introduction of the triangle singularity (TS) mechanism into the axial vector meson decays. While this is a key for a self-consistent description of the axial vector mesons, we will show that by a self-consistent treatment of the TS mechanism there exists smoking-gun observables for disentangling the puzzles in the axial vector spectra and a coherent picture can be obtained.

The TS mechanism in the axial vector decays was recognized with the help of two recent experimental progresses: (I) The observation of abnormally large isospin breaking effects in $J/\psi\to \gamma \eta(1405/1475)\to \gamma +3\pi$ by the BESIII Collaboration~\cite{BESIII:2012aa}, where the $f_1(1420)$ should also contribute. (II) The observation of $a_1(1420)$ in the invariant mass spectrum of $3\pi$ in $\pi^- p\to p+3\pi$ by the COMPASS Collaboration~\cite{Adolph:2015pws}, where the $a_1(1420)$ in the vicinity of $a_1(1260)$ apparently cannot be accommodated by the quark model nonet. For the former case it was first proposed in Ref.~\cite{Wu:2011yx} that the TS mechanism was the key for understanding the abnormally large isospin violations. In a later detailed analysis, it was shown that a small contribution from the $f_1(1420)$ should appear in the decay of $J/\psi\to \gamma +3\pi$~\cite{Wu:2012pg}. Although the importance of the TS mechanism for $\eta(1405/1475)\to 3\pi$ was confirmed by Refs.~\cite{Aceti:2012dj,Achasov:2015uua}, it was emphasized by Ref.~\cite{Achasov:2015uua} that the width effects arising from the intermediate $K^*$ may dilute the contributions from the TS transitions. A comprehensive analysis of the $\eta(1405/1475)$ decays into $K\bar{K}\pi$, $\eta\pi\pi$ and $3\pi$ was presented in Ref.~\cite{Du:2019idk}, and it was clarified that apart from the width effects the TS enhanced $a_0(980)\pi$ production should also be explicitly included. This provides another important isospin breaking source from the TS mechanism in $\eta(1405/1475)\to 3\pi$ and makes it possible to coherently investigate those three decay channels in a self-consistent framework. This has been one of the major motivations of this work since such a treatment can be applied to $J/\psi\to \gamma f_1(1285)/f_1(1420)\to \gamma K\bar{K}\pi$, $\eta\pi\pi$ and $3\pi$, and a better understanding of $f_1(1285)$ and $f_1(1420)$ may be achieved. It should be mentioned that because of the strong enhancement caused by the TS mechanism, it was proposed by Ref.~\cite{Debastiani:2016xgg} that the signal for $f_1(1420)$ observed in $pp$ scatterings by the WA102 Collaboration could be due to the $f_1(1285)$ via the TS mechanism. This is an interesting point since according to the result of Ref.~\cite{Roca:2005nm} the $f_1(1420)$ cannot be accommodated by their model for dynamically generated state.

Concerning the observation of $a_1(1420)$ by the COMPASS Collaboration~\cite{Adolph:2015pws} listed above, 
it is explained to be a tetraquark state in some studies base on QCD sum rule~\cite{Chen:2015fwa,Sundu:2017xct}, and ADS/CFT method~\cite{Gutsche:2017oro}. However, it can be regarded as a natural consequence of the presence of the TS mechanism in this kinematic region caused by the $S$-wave isovector coupling between $K^*\bar{K}+c.c.$ in the $3\pi$ spectrum~\footnote{The TS mechanism was first pointed out by Q.Z. at Hadron 2013 in Nara when the COMPASS data were first reported.}. An analysis based on the TS mechanism was carried out in Ref.~\cite{Ketzer:2015tqa}. By considering that the TS mechanism can be satisfied by the slightly offshell $a_1(1260)$, it gave a natural explanation of the enhancement at about 1.42 GeV in the $3\pi$ channel with $I=1$.  This mechanism was confirmed by  Ref.~\cite{Aceti:2016yeb} where a coupled channel treatment was emphasized. Actually, the $a_1(1420)$ so far has only been observed in $\pi^- p\to p+3\pi$. This can be not only a strong evidence for the TS mechanism, but also a stringent constraint on a coherent analysis. Nevertheless, since the $K^*\bar{K}+c.c.$ can also couple to the negative $C$-parity axial vector states, i.e. $h_1$, $h_1'$ and $b_1$, examinations of its impact on the productions and decays of these states can further well-establish the axial vector meson spectra.

To proceed, In Section II we first give a brief introduction of the axial mixing scenarios, which are closely correlated with the TS mechanism due to the strong couplings of the axial vector states to vector and pseudoscalar mesons. We then make a detailed analysis of the productions and decays of the two sets of axial vector mesons with the presence of the TS mechanism in Section III. The numerical results and discussions are presented in Section IV, and a brief summary is given in Section V.

\section{State mixings and determination of the mixing angles}

In the axial vector sector, the mixing angles between $f_1(1285)$ and $f_1(1420)$, and between $h_1(1170)$ and $h_1(1415)$, are correlated with the mixing between $K_1(1270)$ and $K_1(1400)$ in the quark model. Since the mixing angles decide the relative coupling strengths of the two states in a doublet ($f_1(1285)/f_1(1420)$ or $h_1(1170)/h_1(1415)$) to $K^*\bar{K}+c.c.$,  we first set up the convention for the mixing schemes, and discuss the leading-order couplings in the flavor SU(3) symmetry limit. These couplings will be the input for evaluating the relative production rates for the axial vector meson productions in $J/\psi$ decays and the impact of the TS mechanism in these hadronic decays. For simplicity, from now on, we denote the $f_1(1285)$, $f_1(1420)$, $h_1(1170)$ and $h_1(1415)$ by $f$, $f'$, $h$ and $h'$, respectively.

We start with the state mixing on the SU(3) basis,
\begin{eqnarray}
\left(
  \begin{array}{c}
    f_1' \\
    f_1 \\
  \end{array}
\right)
&=&
\left(
  \begin{array}{cc}
     \cos{\theta_f} & -\sin{\theta_f}\\
     \sin{\theta_f} & \cos{\theta_f}
  \end{array}
\right)
\left(
  \begin{array}{c}
    \tilde{f}_8 \\
    \tilde{f}_1 \\
  \end{array}
\right)=
\left(
  \begin{array}{cc}
     \cos{\theta_f} & -\sin{\theta_f}\\
     \sin{\theta_f} & \cos{\theta_f}
  \end{array}
\right)
\left(
  \begin{array}{cc}
     \sqrt{\frac{1}{3}} & -\sqrt{\frac{2}{3}}\\
     \sqrt{\frac{2}{3}} & \sqrt{\frac{1}{3}}
  \end{array}
\right)
\left(
  \begin{array}{c}
    f_n \\
    f_s \\
  \end{array}
\right)\nonumber\\
&=&
\left(
  \begin{array}{cc}
     \cos{\alpha_f} & -\sin{\alpha_f}\\
     \sin{\alpha_f} & \cos{\alpha_f}
  \end{array}
\right)
\left(
  \begin{array}{c}
    f_n \\
    f_s \\
  \end{array}
\right)
,
\end{eqnarray}
where $\tilde{f}_1\equiv (u\bar{u}+d\bar{d}+s\bar{s})/{\sqrt{3}}$, $\tilde{f}_8\equiv (u\bar{u}+d\bar{d}-2s\bar{s})/{\sqrt{6}}$, $f_n\equiv (u\bar{u}+d\bar{d})/{\sqrt{2}}$ and $f_s\equiv s\bar{s}$. Likewise, the mixing angles between $h$ and $h'$ states are defined by
\begin{eqnarray}
\left(
  \begin{array}{c}
    h_1' \\
    h_1 \\
  \end{array}
\right)
&=&
\left(
  \begin{array}{cc}
     \cos{\theta_h} & -\sin{\theta_h}\\
     \sin{\theta_h} & \cos{\theta_h}
  \end{array}
\right)
\left(
  \begin{array}{c}
    \tilde{h}_8 \\
    \tilde{h}_1 \\
  \end{array}
\right)=
\left(
  \begin{array}{cc}
     \cos{\theta_h} & -\sin{\theta_h}\\
     \sin{\theta_h} & \cos{\theta_h}
  \end{array}
\right)
\left(
  \begin{array}{cc}
     \sqrt{\frac{1}{3}} & -\sqrt{\frac{2}{3}}\\
     \sqrt{\frac{2}{3}} & \sqrt{\frac{1}{3}}
  \end{array}
\right)
\left(
  \begin{array}{c}
    h_n \\
    h_s \\
  \end{array}
\right)\nonumber\\
&=&
\left(
  \begin{array}{cc}
     \cos{\alpha_h} & -\sin{\alpha_h}\\
     \sin{\alpha_h} & \cos{\alpha_h}
  \end{array}
\right)
\left(
  \begin{array}{c}
    h_n \\
    h_s \\
  \end{array}
\right),\label{mixingangleoff1}
\end{eqnarray}
where $\tilde{h}_1\equiv (u\bar{u}+d\bar{d}+s\bar{s})/{\sqrt{3}}$, $\tilde{h}_8\equiv(u\bar{u}+d\bar{d}-2s\bar{s})/{\sqrt{6}}$, $h_n\equiv (u\bar{u}+d\bar{d})/{\sqrt{2}}$ and $h_s\equiv s\bar{s}$. The relation between $\theta_f$ ($\theta_h$) and $\alpha_f$ ($\alpha_h$) is
\begin{eqnarray}
\alpha_{f(h)}=\theta_{f(h)}+\arctan{\sqrt{2}} \  .
%\ \alpha_h=\theta_h+\arctan{\sqrt{2}}.
\end{eqnarray}
The mixing angles $\theta_{f(h)}$ encode the mechanisms that contribute to the mass matrices via the mixings between the SU(3) flavor singlet and octet, i.e. $\tilde{f}_1$ and $\tilde{f}_8$ ($\tilde{h}_1$ and $\tilde{h}_8$).

With the help of the Gell-Mann-Okubo relations (see Appendix for a pedagogic deduction), the octet masses $m_{\tilde{f}_8}^2$ and $m_{\tilde{h}_8}^2$ can be expressed by the states with isospin 1/2 and 1 in the quark model, respectively,
i.e.
\begin{eqnarray}
m_{\tilde{f}_8}^2&=&\frac{4m_{K_{1A}}^2-m_{a_1}^2}{3},\nonumber\\
m_{\tilde{h}_8}^2&=&\frac{4m_{K_{1B}}^2-m_{b_1}^2}{3},
\end{eqnarray}
where $K_{1A}$ and $K_{1B}$ are assigned as the $^3P_1$ and $^1P_1$ states, respectively. The mixing angles, $\theta_f$ and $\theta_h$, can thus be calculated by
\begin{eqnarray}
\tan{\theta_f}&=&\frac{\sqrt{m_{\tilde{f}_8}^2(m_f^2+m_{f'}^2-m_{\tilde{f}_8}^2)-m_f^2m_{f'}^2}}{m_{\tilde{f}_8}^2-m_f^2}=\frac{m_{f'}^2-m_{\tilde{f}_8}^2}{\sqrt{m_{\tilde{f}_8}^2(m_f^2+m_{f'}^2-m_{\tilde{f}_8}^2)-m_f^2m_{f'}^2}},\\
\tan{\theta_h}&=&\frac{\sqrt{m_{\tilde{h}_8}^2(m_h^2+m_{h'}^2-m_{\tilde{h}_8}^2)-m_h^2m_{h'}^2}}{m_{\tilde{h}_8}^2-m_h^2}=\frac{m_{h'}^2-m_{\tilde{h}_8}^2}{\sqrt{m_{\tilde{h}_8}^2(m_h^2+m_{h'}^2-m_{\tilde{h}_8}^2)-m_h^2m_{h'}^2}}.
\label{tantheta}
\end{eqnarray}
One can see from the above relation that given the masses of the physical states and the octet masses $m_{\tilde{f}_8}^2$ and $m_{\tilde{h}_8}^2$, the mixing angles can be determined. Note that the octet masses $m_{\tilde{f}_8}^2$ and $m_{\tilde{h}_8}^2$ are determined by the masses of $K_{1A}$ and $K_{1B}$ which, however, are not the physical masses. The determination of $\theta_{f(h)}$ is thus correlated with the mixing between $K_{1A}$ and $K_{1B}$.

The corresponding physical states, $K_1(1270)$ and $K_1(1400)$, as the mixing states between $K_{1A}$ and $K_{1B}$ (i.e. $^3P_1$ and $^1P_1$) are expressed as
\begin{eqnarray}
\left(
  \begin{array}{c}
    K_1(1270) \\
    K_1(1400) \\
  \end{array}
\right)
=
\left(\begin{array}{cc}
\cos{\theta_{K_1}} & \sin{\theta_{K_1}}\\
-\sin{\theta_{K_1}} & \cos{\theta_{K_1}}
\end{array}
\right)
\left(
  \begin{array}{c}
    K_{1B} \\
    K_{1A} \\
  \end{array}
\right).
\end{eqnarray}
As a result, the mixing angles $\theta_{f(h)}$ are now correlated with the dynamics for the $K_{1A}$ and $K_{1B}$ mixing via their mixing angle $\theta_{K_1}$.  

There have been various approaches for the determination of $\theta_{K_1}$ in the literature. The early analysis of $\tau$ decay in Ref.~\cite{Suzuki:1993yc} gave a two-fold solution, i.e. $|\theta_{K_1}|=33^{\circ}$ or $57^{\circ}$. With the help of the constituent quark model, Ref.~\cite{Blundell:1995au} predicted $|\theta_{K_1}|\sim 45^{\circ}$. In Ref.~\cite{Burakovsky:1997dd} a range of $35^{\circ}\le|\theta_{K_1}|\le 55^{\circ}$ was obtained in the nonrelativistic quark model. In Ref.~\cite{Cheng:2003bn} $\theta_{K_1}\sim-58^{\circ}$ was shown to be favored in the charmed meson decays. Analysis of the axial vector decays into a pseudoscalar and a vector meson suggested $\theta_{K_1}=+(62\pm3)^{\circ}$~\cite{Roca:2003uk}, which seems to be confirmed by the conclusion in Ref.~\cite{Li:2006we}. Later, based on the data for the $B$ and $f_1$ radiative decays, $\theta_{K_1}$ was determined to be $-34\pm13^{\circ}$ by Ref.~\cite{Yang:2010ah}, together with $\theta_f=(19.4^{+4.5}_{-4.6})^{\circ}$. The analysis of Ref.~\cite{Cheng:2011pb} also favored $|\theta_{K_1}|\sim35^{\circ}$. The sign of $\theta_{K_1}$ was discussed in Ref.~\cite{Dag:2012zz}, which led to $\theta_{K_1}=-(39\pm4)^{\circ}$. A combined analysis including all light axial vector mesons was made in Ref.~\cite{Divotgey:2013jba}, where a good agreement with the experimental data was achieved with $\theta_{K_1}=-(33.6\pm4.3)^{\circ}$. In Ref.~\cite{Cheng:2013cwa} a brief review of the status of $\theta_{K_1}$ suggested that $\theta_{K_1}\sim33^{\circ}$ (less than $\pi/4$) is more favoured than $57^{\circ}$ (larger than $\pi/4$). Note that the sign ambiguity was removed by fixing the relative sign of the coupling constants for $K_{1A}$ and $K_{1B}$~\cite{Cheng:2013cwa}.
The recent analysis of $B^+\to J/\psi K_1^+$ using a pQCD method also obtained $\theta_{K_1}\sim 33^{\circ}$~\cite{Zhang:2017cbi}.

Taking the PDG~\cite{Patrignani:2016xqp} values for these charge neutral axial vector states (see Table~\ref{axial-mass}) and adopting several possible solutions for $\theta_{K_1}$ as inputs, we extract the mixing angles $\theta_{f(h)}$ and $\alpha_{f(h)}$ ($\alpha_{f(h)}=\theta_{f(h)}+\arctan{\sqrt{2}}$) in Table~\ref{sampleangle}. One can see that the large range of values for $\theta_{K_1}$ leads to also large uncertainties in the predictions of $\theta_{f(h)}$ and $\alpha_{f(h)}$. 

\begin{table}
\caption{Physical masses adopted for the axial vector mesons~\cite{Patrignani:2016xqp}.}\label{axial-mass}
\centering
\begin{tabular}{c|c}
  \hline\hline
 Mesons & Mass (GeV) \\
 \hline
 $a_1$ & 1.26 \\
 $f_1$ & 1.285 \\
 $f_1'$ & 1.426 \\
 $b_1$ & 1.235 \\
 $h_1$ & 1.17 \\
 $h_1'$ & 1.423 \\
 \hline
\end{tabular}
\end{table}

\begin{table}
\caption{Extracted mixing angles for $\theta_{f(h)}$ and $\alpha_{f(h)}$ with typical values for $|\theta_{K_1}|$ as the input.}\label{sampleangle}
\centering
\begin{tabular}{cccccc}
  \hline\hline
  $|\theta_{K_1}|$ & $33^{\circ}$ & $34^{\circ}$ & $39^{\circ}$ & $45^{\circ}$ & $57^{\circ}$ \\
  \hline
  $\alpha_f-90^{\circ}$ & $-6.84^{\circ}$ & $-5.52^{\circ}$ & $0.82^{\circ}$ & $8.17^{\circ}$ & $22.7^{\circ}$  \\
  \hline
  $\alpha_h-90^{\circ}$ & $2.45^{\circ}$ & $1.77^{\circ}$ & $-1.80^{\circ}$& $-6.55^{\circ}$  & $-18.1^{\circ}$  \\
  \hline
  $\theta_f$ & $28.4^{\circ}$ & $29.7^{\circ}$ & $36.1^{\circ}$ & $43.4^{\circ}$ & $58.0^{\circ}$  \\
  \hline
  $\theta_h$ & $37.7^{\circ}$ & $37.0^{\circ}$ & $33.5^{\circ}$& $28.7^{\circ}$  & $17.2^{\circ}$  \\
  \hline\hline
\end{tabular}
\end{table}

In order to determine $\theta_{f(h)}$ ($\alpha_{f(h)}$) we look for alternative constraints on the mixing angles. It seems that the $h_1$ and $h_1'$ decays into a vector and a pseudoscalar meson can set up a reasonable constraint.

The Lagrangian describing the vertices between the $1^{+-}$ axial vector meson (B), vector meson (V) and pseudoscalar meson (P) is given by
\begin{eqnarray}\label{Lag-bvp}
L_{BVP}&=&g_{BVP}Tr[B^{\mu}\{V_{\mu},P\}],
\end{eqnarray}
where $g_{BVP}$ is the coupling constant and with the SU(3) flavor symmetry the coupling fields are
\begin{eqnarray}
B^{\mu}&\equiv&\left(
\begin{array}{ccc}
 \frac{\cos{\alpha_h}h_1'+\sin{\alpha_h}h_1}{\sqrt{2}}+\frac{b_1^0}{\sqrt{2}} & b_1^+ & K_{1B}^+\\
  b_1^- & \frac{\cos{\alpha_h}h_1'+\sin{\alpha_h}h_1}{\sqrt{2}}-\frac{b_1^0}{\sqrt{2}} & K_{1B}^0 \\
  K_{1B}^- & \bar{K}_{1B}^0 & -\sin{\alpha_h}h_1'+\cos{\alpha_h}h_1
\end{array}
\right),
\end{eqnarray}
\begin{eqnarray}
V^{\mu}&\equiv &\left(
\begin{array}{ccc}
  \frac{\rho^0}{\sqrt{2}}+\frac{\omega}{\sqrt{2}} & \rho^+ & K^{*+} \\
  \rho^- & \frac{-\rho^0}{\sqrt{2}}+\frac{\omega}{\sqrt{2}} & K^{*0} \\
  K^{*-} & \bar{K}^{*0} & \phi
\end{array}
\right), \\
P&\equiv &\left(
\begin{array}{ccc}
  \frac{\pi^0}{\sqrt{2}}+\frac{\cos{\alpha_P}\eta+\sin{\alpha_P}\eta'}{\sqrt{2}} & \pi^+ & K^+ \\
  \pi^- & -\frac{\pi^0}{\sqrt{2}}+\frac{\cos{\alpha_P}\eta+\sin{\alpha_P}\eta'}{\sqrt{2}} & K^0 \\
  K^- & \bar{K}^0 & \sin{\alpha_P}\eta+\cos{\alpha_P}\eta'
\end{array}
\right),
\end{eqnarray}
where we parametrize the mixing between $\eta$ and $\eta'$ as
\begin{eqnarray}\label{pseudoscalar-mixing}
\left(
  \begin{array}{c}
    \eta \\
    \eta' \\
  \end{array}
\right)
=
\left(\begin{array}{cc}
\cos{\alpha_P} & -\sin{\alpha_P}\\
\sin{\alpha_P} & \cos{\alpha_P}
\end{array}
\right)
\left(
  \begin{array}{c}
    \eta_n \\
    \eta_s \\
  \end{array}
\right).
\end{eqnarray}

The couplings for an vector meson to two pseudoscalars have the following form:
\begin{eqnarray}
L_{VPP}&=& ig_{VPP}Tr[V^{\mu}(\partial_{\mu}PP-P\partial_\mu P)],
\end{eqnarray}
where $g_{VPP}$ is the coupling constant and can be calculated by $K^*\to K\pi$ or $\phi\to K\bar{K}$. With the data for $\phi\to K\bar{K}$~\cite{Patrignani:2016xqp}, one determines $g_{VPP}=4.52$. To determine the coupling $g_{BVP}$, we assume that the total width of $h_1(1170)$ is saturated by $h_1\to\rho\pi\to\pi^+\pi^-\pi^0$ for which the amplitude  reads
\begin{eqnarray}
M_{h_1\to\rho\pi\to\pi^+\pi^-\pi^0}&=&
ig_{h_1\rho\pi}i^2g_{\rho\pi\pi}\epsilon_{\mu}\frac{i}{s_{cd}-m_{\rho}^2+im_{\rho}\Gamma_{\rho}}i[(1+\frac{s_d-s_a}{s_{cd}})p_a^{\mu}+(-1+\frac{s_d-s_a}{s_{ad}})p_d^{\mu}]\nonumber\\
&+& ig_{h_1\rho\pi}i^2g_{\rho\pi\pi}\epsilon_{\mu}\frac{i}{s_{ab}-m_{\rho}^2+im_{\rho}\Gamma_{\rho}}i[(-1+\frac{s_a-s_b}{s_{ab}})p_a^{\mu}+(1+\frac{s_a-s_b}{s_{ab}})p_b^{\mu}]\nonumber\\
&+& ig_{h_1\rho\pi}i^2g_{\rho\pi\pi}\epsilon_{\mu}\frac{i}{s_{bd}-m_{\rho}^2+im_{\rho}\Gamma_{\rho}}i[(-1+\frac{s_b-s_d}{s_{bd}})p_b^{\mu}+(1+\frac{s_b-s_d}{s_{bd}})p_d^{\mu}],
\label{hhtorhopi}
\end{eqnarray}
where the momenta of $\pi^+$, $\pi^0$, $\rho$ and $\pi^-$ mesons are denoted as $p_a$, $p_b$, $p_c$ and $p_d$, respectively, and $s_{ad}\equiv (p_a+p_d)^2$, $s_{ab}\equiv (p_a+p_b)^2$, $s_{bd}\equiv (p_b+p_d)^2=s_1+s_b-s_{ad}+s_a+s_d-s_{ab}$. Given $\Gamma_{h_1\to\rho\pi\to\pi^+\pi^-\pi^0}=0.37$ GeV from the PDG as input, one extracts $g_{h_1\rho\pi}=4.28$ GeV. 

Note that the following relation holds,
\begin{equation}
g_{h_1\rho\pi}\equiv \sqrt{2}g_{BVP}\sin{\alpha_h}.
\end{equation}  
It suggests that another relation is needed in order to determine $g_{BVP}$ and $\alpha_h$ in the $1^{+-}$ sector. In principle, with the data for $h_1'\to\rho\pi\to\pi^+\pi^-\pi^0$, one will be able to determine these two quantities. However, the branching ratio of $h_1'\to\rho\pi\to\pi^+\pi^-\pi^0$ has not been well established. To get around of this problem, it shows that the total width measurement may provide a reliable estimate given that it is dominated ( or nearly saturated) by the $\rho\pi$ and $K^*\bar{K}$ channels. Since these two channels can be related to each other by the SU(3) flavor symmetry, one has 
\begin{eqnarray}
g_{h_1'\rho\pi}&=&\sqrt{2}g_{BVP}\cos{\alpha_h}=g_{h_1\rho\pi}\cot{\alpha_h} \ , \\
g_{h_1'K^*\bar{K}}&=&g_{BVP}\left(\frac{\sqrt{2}}{2}\cos{\alpha_h}-\sin{\alpha_h}\right) \\
&=& \frac{1}{2} g_{h_1\rho\pi}(\cot{\alpha_h}-\sqrt{2}) \ .
\end{eqnarray}
Then, the total width of $h_1'$ will be calculated as the sum of the partial widths of the $\rho\pi$ and $K^*\bar{K}$ channels. 

The BESIII Collaboration recently measured the total width of $h_1'$, i.e. $\Gamma_{h'}=90\pm 9.8\pm 17.5$ MeV~\cite{Ablikim:2018ctf}. We find that with $\theta_{K_1}=34^{\circ}$ (i.e. $\alpha_h=91.77^\circ$) and $g_{h_1\rho\pi}=4.28$ GeV, the parameter relations give $g_{BVP}=3.03$ GeV and $g_{h_1'K^*\bar{K}}=-3.09$ GeV. Thus, the partial width of $h_1'\to K^*\bar{K}\to K\bar{K}\pi$ is estimated to be
\begin{equation}
\Gamma_{h_1'\to K^*\bar{K}\to K\bar{K}\pi}\simeq 6\Gamma_{h_1'\to K^*\bar{K}\to K^+K^-\pi^0}=55.9 \ \text{MeV}.\label{kkpiwidthofh1h}
\end{equation}
Together with the contributions from the other possible decay channels, the total width of $h_1'$ can be reasonably described. We also mention that with higher values for $|\theta_{K_1}|$ in Table~\ref{sampleangle}, e.g. $|\theta_{K_1}|=57^{\circ}$, the partial width for  $h_1'\to\rho\pi\to\pi^+\pi^-\pi^0$ will be large and in contradiction with the experimental measurements. So, we adopt $\theta_{K_1}=34^{\circ}$ as the input for the later calculations.

\section{Production and decay of axial vector mesons with $J^{PC}=1^{++}$}

\subsection{Key issues }

In our previous study~\cite{Wu:2012pg} it was shown that the axial vector meson $f_1'$ should have contributions to the isospin-violating decay of $J/\psi\to \gamma+ 3\pi$ which was measured by BESIII~\cite{BESIII:2012aa}. The observable of the angular distribution of the pion recoiling the $f_0(980)$ should contain a non-negligible $S$-wave contribution apart from the dominant $P$-wave from the intermediate $\eta$ resonances. Different from the treatment of Ref.~\cite{Wu:2012pg} where an arbitrary coupling strength was determined by fitting the angular distribution, we will quantify the production and decay of both $f_1$ and $f_1'$ in the $J/\psi$ radiative decays into $\gamma\eta\pi\pi$, $\gamma K\bar{K}\pi$ and $\gamma + 3\pi$ with the presence of the TS mechanism. 

In Ref.~\cite{APFB-2017} a preliminary result was reported and it showed that the $f_1(1420)$ as a pole structure should account for the $S$-wave enhancements observed in the invariant mass spectra of $\eta\pi\pi$ and $K\bar{K}\pi$ in $J/\psi\to \eta\pi\pi$ and $K\bar{K}\pi$, respectively. Later, it was claimed in Ref.~\cite{Debastiani:2016xgg} that the $f_1(1420)$ enhancement was not a genuine state and it could be produced by the TS mechanism due to the $f_1(1285)$ pole. This is an interesting scenario since it, on the one hand, shows that the TS mechanism plays an non-negligible role in various processes indeed. And on the other hand, it also raises questions on the structure of $f_1(1420)$. In Refs.~\cite{Roca:2005nm,Roca:2006tr} a systematic study of the $S$-wave vector and peudoscalar meson interactions led to dynamically generated pole structures which can be associated with the PDG listed states, i.e. $b_1(1235), \ h_1(1170), \ h_1(1380), \ a_1(1260)$, and $f_1(1285)$. It was also mentioned there that the $f_1(1420)$ could not be accommodated in their scheme. In Ref.~\cite{Debastiani:2016xgg} the $f_1(1420)$ was proposed to be an enhanced structure by the TS mechanism which was originated from the $f_1(1285)$ as a dynamically generated state by the $K^*\bar{K}+c.c.$ $S$-wave interaction. The authors of Ref.~\cite{Debastiani:2016xgg} obtained the lineshape of the invariant mass by fitting the data from WA102~\cite{Barberis:1998by} with an arbitrary strength undetermined. Although this seems to be consistent with the result of Ref.~\cite{Roca:2005nm}, it would leave crucial questions in the understanding the overall axial vector spectrum. In particular, by treating $f_1(1285)$ as a dynamically generated state due to the $S$-wave $K^*\bar{K}+c.c.$ interaction would bring conflicts to some of those experimental observables for $f_1'$ which fit quite well the flavor singlet and octet mixing pattern~\cite{Barberis:1998by,Close:1997nm,Patrignani:2016xqp}.

To proceed, we first point out that the dominant decay mode of $f_1(1285)$ is $\eta\pi\pi$ with $B.R.(f_1\to \eta\pi\pi)=(52.4^{+1.9}_{-2.2})\%$, among which $B.R.(f_1\to a_0(980)\pi)=(36\pm 7)\%$~\cite{Patrignani:2016xqp}. Note that the branching ratio $B.R.(f_1\to a_0(980)\pi)=(36\pm 7)\%$ does not include the contributions from the $a_0(980)\to K\bar{K}$~\cite{Patrignani:2016xqp}. It means that the real value should be larger than this and it explicitly indicates the dominant mode of $f_1\to a_0(980)\pi$. In contrast, the branching ratio for $f_1\to K\bar{K}\pi$ is $B.R.(f_1\to K\bar{K}\pi)=(9.0\pm 0.4)\%$ and the intermediate decay mode of $f_1\to K^*\bar{K}+c.c.$ has not been observed. Although the phase space would limit the partial width of $f_1\to K^*\bar{K}+c.c.$, 
it suggests that the $f_1$ coupling to $K^*\bar{K}$ does not necessarily to be strong. This actually jeopardizes the conclusion of Ref.~\cite{Debastiani:2016xgg} where it has been assumed that the $K^*\bar{K}+c.c.$ decay channel had saturated the $K\bar{K}\pi$ mode for $f_1(1285)$. Another caveat in Ref.~\cite{Debastiani:2016xgg} is that the invariant mass spectrum measured by WA102 cannot constrain the relative coupling strength for the production of $f_1$ and $f_1'$ and this will be correlated with the couplings for their decays into exclusive channels (Note that their mixing angle is also a variable to be determined). These remind us of a coherent study of both $f_1$ and $f_1'$ in the exclusive process of the $J/\psi$ radiative decays where both the production and decay mechanisms can be properly evaluated.

For the coherent study of the axial vector mesons with $C=1^{++}$ the following observations can be itemized:
\begin{itemize}

\item The available experimental results suggest that the $f_1'$ dominantly decays into the $K\bar{K}\pi$ final state via $K^*\bar{K}+c.c.$ and its decays into $\eta\pi\pi$ is much suppressed.

\item The $f_1$ dominantly decays into $\eta\pi\pi$ via the intermediate $a_0\pi$ process and its decay into $K\bar{K}\pi$ is also via the $a_0\pi$.

\item The mixing between $f_1$ and $f_1'$ will affect both production and decay processes.

\item The $S$-wave coupling to $K^*\bar{K}+c.c.$ introduces the effects from the TS mechanism which needs to be coherently investigated for both $f_1$ and $f_1'$ in exclusive processes.

\item The $S$-wave coupling to $K^*\bar{K}+c.c.$ implies the TS enhancement in the isospin-1 channel which refers to the $a_1(1420)$.

\end{itemize}

\subsection{Productions of the $C=+1$ axial vector mesons in charmonium decays}

The isoscalar $f_1$ and $f_1'$ can be produced in the $J/\psi$ radiative decays. In their three-body decays, e.g. $f_1\to \eta\pi\pi$ and $K\bar{K}\pi$, both states can contribute. It thus requires a coherent study of these two states in each channel. We parametrize the production amplitudes using the following effective Lagrangian for $J/\psi\to\gamma A$ (here, $A$ denotes $f_1$ or $f_1'$):
\begin{eqnarray}\label{Ljpsigammaf1}
L_{J/\psi\gamma A}=g_A\epsilon_{\mu\nu\rho\sigma}\partial^{\mu}\psi^{\nu}\gamma^{\rho}A ^{\sigma} \ ,
\end{eqnarray}
which gives the amplitude
\begin{eqnarray}
M_{J/\psi\gamma A}=g_A\epsilon_{\mu\nu\rho\sigma}p_{\psi}^{\mu}\epsilon_{\psi}^{\nu}\epsilon_{\gamma}^{\rho}\epsilon_{A}^{\sigma} \ ,
\end{eqnarray}
where $\epsilon_{\psi}^{\nu}$, $\epsilon_{\gamma}^{\rho}$, and $\epsilon_{A}^{\sigma}$ are the polarization vectors for $J/\psi$, the photon and the axial vector meson, respectively. The production strength for $f_1$ and $f_1'$ in the $J/\psi$ radiative decays can be related to each other in the SU(3) flavor symmetry limit:
\begin{eqnarray}\label{prod-ratio}
\frac{g_f'}{g_f}=\frac{\langle f_1'|\hat{H}_\gamma|J/\psi\rangle}{\langle f_1|\hat{H}_\gamma|J/\psi\rangle}=\frac{\sqrt{2}\cos{\alpha_f}-\sin{\alpha_f}}{\sqrt{2}\sin{\alpha_f}+\cos{\alpha_f}},
\end{eqnarray}
where we have assumed that $\langle (s\bar{s})_{1^{++}}|\hat{H}_\gamma|J/\psi\rangle=\langle (u\bar{u})_{1^{++}}|\hat{H}_\gamma|J/\psi\rangle=\langle (d\bar{d})_{1^{++}}|\hat{H}_\gamma|J/\psi\rangle$. Namely, we neglect the SU(3) symmetry breaking with the couplings for the light pair creations. Then, the amplitude of $J/\psi\to\gamma (f_1+f_1')\to\gamma ABC$ can be parametrized as
\begin{eqnarray}
M_{J/\psi\to\gamma (f_1+f_1')\to\gamma ABC}=M_{J/\psi\gamma f_1}\left(\frac{i}{D_{f_1}}M_{f_1\to ABC}+\frac{g_f'}{g_f}\frac{i}{D_{f_1'}}M_{f_1'\to ABC}\right) ,
\end{eqnarray}
where $M_{f_1\to ABC}$ and $M_{f_1'\to ABC}$ are the decay amplitudes for $f_1$ and $f_1'$, respectively, and will be explicitly calculated in the next subsection.

For the convenience of calculation, we define $M_{f_1'\to ABC}=\sum_{i}M_{f_1'\to ABC}^{(i)}$ and $M_{f_1'\to ABC}^{(i)}=\chi_i M_{f_1\to ABC}^{(i)}$, where $i$ denotes the $i$-th intermediate state. Then, the spectrum can be expressed as the following form: 
\begin{eqnarray}
\frac{d\Gamma}{d\sqrt{s}}=\frac{2s}{\pi}\Gamma_{J/\psi\to\gamma f_1}\frac{1}{2\sqrt{s}}\int d\Phi_{ABC}\frac{1}{3}|\sum_i\left(\frac{i}{D_{f_1}}+\frac{\chi_ig_f'}{g_f}\frac{i}{D_{f_1'}}\right)M_{f_1\to ABC}^{(i)}|^2,
\end{eqnarray}
where the $\Phi_{ABC}$ is the phase space for the three-body $ABC$ in $J/\psi\to \gamma ABC$; Functions $D_{f_1}$ and $D_{f_1'}$ are the inverse propagators of $f_1$ and $f_1'$, respectively, i.e.
\begin{eqnarray}
D_{f_1}&=&s-m_{f_1}^2+im_{f_1}\Gamma_{f_1}, \\
D_{f_1'}&=&s-m_{f_1'}^2+im_{f_1'}\Gamma_{f_1'}(s),
\end{eqnarray}
where $\Gamma_{f_1}=22.7 \ \text{MeV}$ and $\Gamma_{f_1'}=\Gamma_{f_1'\to K\bar{K}\pi}+\Gamma_{f_1'\to a_0\pi\to\eta\pi\pi}$, are adopted.

Different from the production of $f_1/f_1'$ the production of $a_1$  in the $J/\psi$ radiative decays will be suppressed due to the isospin.  To produce $a_1$ in the $J/\psi$ radiative decays it needs to go through the final state photon radiation. This is relatively suppressed by the additional gluon exchange in $c\bar{c}$ annihilations. In $e^+e^-$ annihilations the ideal process to probe the production of $a_1$ is via $\chi_{c1}\to \pi a_1(1260)/a_1(1420)\to 4\pi$, where $\chi_{c1}$ can be copiously produced through $\psi(3686)\to \gamma\chi_{c1}$. For the production of $\chi_{c1}\to \pi a_1$ it can be described by the following Lagrangian:
\begin{equation}
L_{\chi AP}=i g_a\epsilon_{\mu\nu\rho\sigma}\partial^{\mu}\chi_{c1}^{\nu}\partial^{\rho}A^{\sigma} P \ ,
\end{equation}
where $g_a$ is the coupling strength for producing a pair of $(q\bar{q})_{1^{++}}$ and $(q\bar{q})_{0^{-+}}$ in the $\chi_{c1}$ decays. 

In principle, $g_a$ can be determined by the partial decay width for $\chi_{c1}\to \pi a_1(1260)$. Note that this quantity is correlated with the measurement of the exclusive process with the $a_1$ decays into final stable particles, e.g. $a_1\to 3\pi$. At this moment, the data are not available. However, for the purpose of determining the lineshape as a characteristic feature $g_a$ can be treated as an overall parameter for the production strength. So we need only consider the $3\pi$ spectrum in the analysis which can be described as  
\begin{eqnarray}
\frac{d\Gamma}{d\sqrt{s}}\sim\frac{2s}{\pi}\frac{\Gamma_{a_1\to ABC}(s)}{(s-m_{a_1(1260)})^2+s\Gamma_{a_1}(s)},
\end{eqnarray}
where the $\Gamma_{a_1\to ABC}$ represents the energy-dependent width for $a_1^-\to \rho^0\pi^-\to\pi^+\pi^-\pi^0$ or $a_1^-\to f_0\pi^-\to\pi^+\pi^-\pi^0$. 
The energy-dependent total width $\Gamma_{a_1}(s)$ is estimated by 
\begin{equation}
\Gamma_{a_1}(s)\simeq \Gamma_{a_1\to \rho\pi \ (S-\text{wave})}/B.R.(a_1\to\rho\pi \ (S-\text{wave})) \ .
\end{equation}
Note that both the total width and partial decay width $\Gamma_{a_1\to \rho\pi \ (S-\text{wave})}$ are far from well established for $a_1(1260)$~\cite{Patrignani:2016xqp}, although the early measurement by the CLEO Collaboration suggested $B.R.(a_1\to\rho\pi \ (S-\text{wave}))\simeq 0.6$~\cite{Asner:1999kj}. This will prevent us from extracting the absolute value for the partial width for $\chi_{c1}\to \pi a_1(1260)$, but will not affect the study of the lineshape since both the total width and partial decay width are not expected to be sensitive to the invariant mass spectrum of the $3\pi$ in the range of 1.2$\sim$ 1.4 GeV. Moreover, one notices that the coupling for $a_1\to\rho\pi$ can be related to that for $a_1\to  K^*\bar{K}$ with the SU(3) flavor symmetry. This allows us to estimate the relative significance of the TS mechanism in the $3\pi$ spectrum in comparison with the tree process of $a_1\to\rho\pi\to 3\pi$. This will be shown in the study of $a_1\to 3\pi$ later.

\subsection{Decay mechanisms for $f_1$ and $f_1'$ into $\eta\pi\pi$, $K\bar{K}\pi$, and $3\pi$}

\begin{figure}
  \centering
  \includegraphics[width=2.6in]{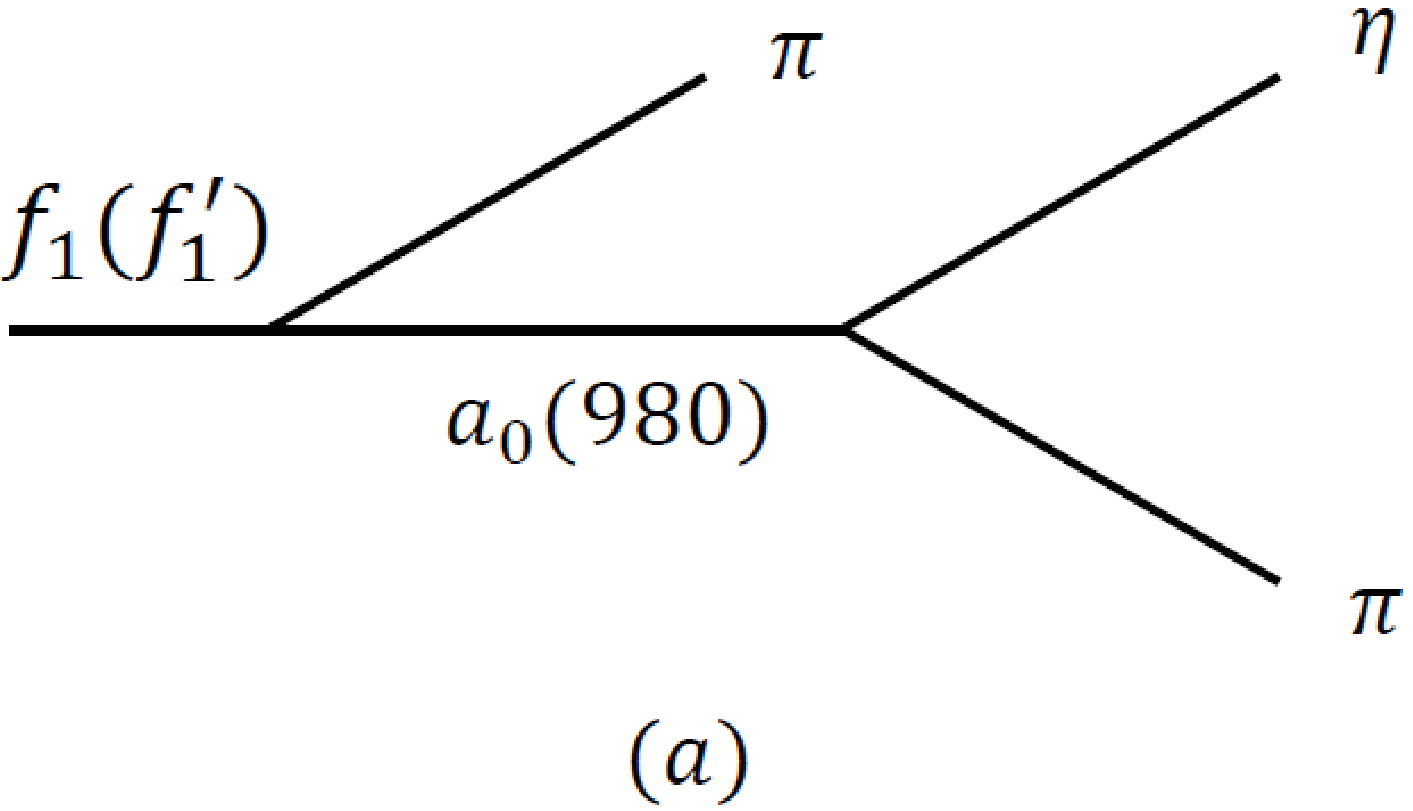}
  \includegraphics[width=2.6in]{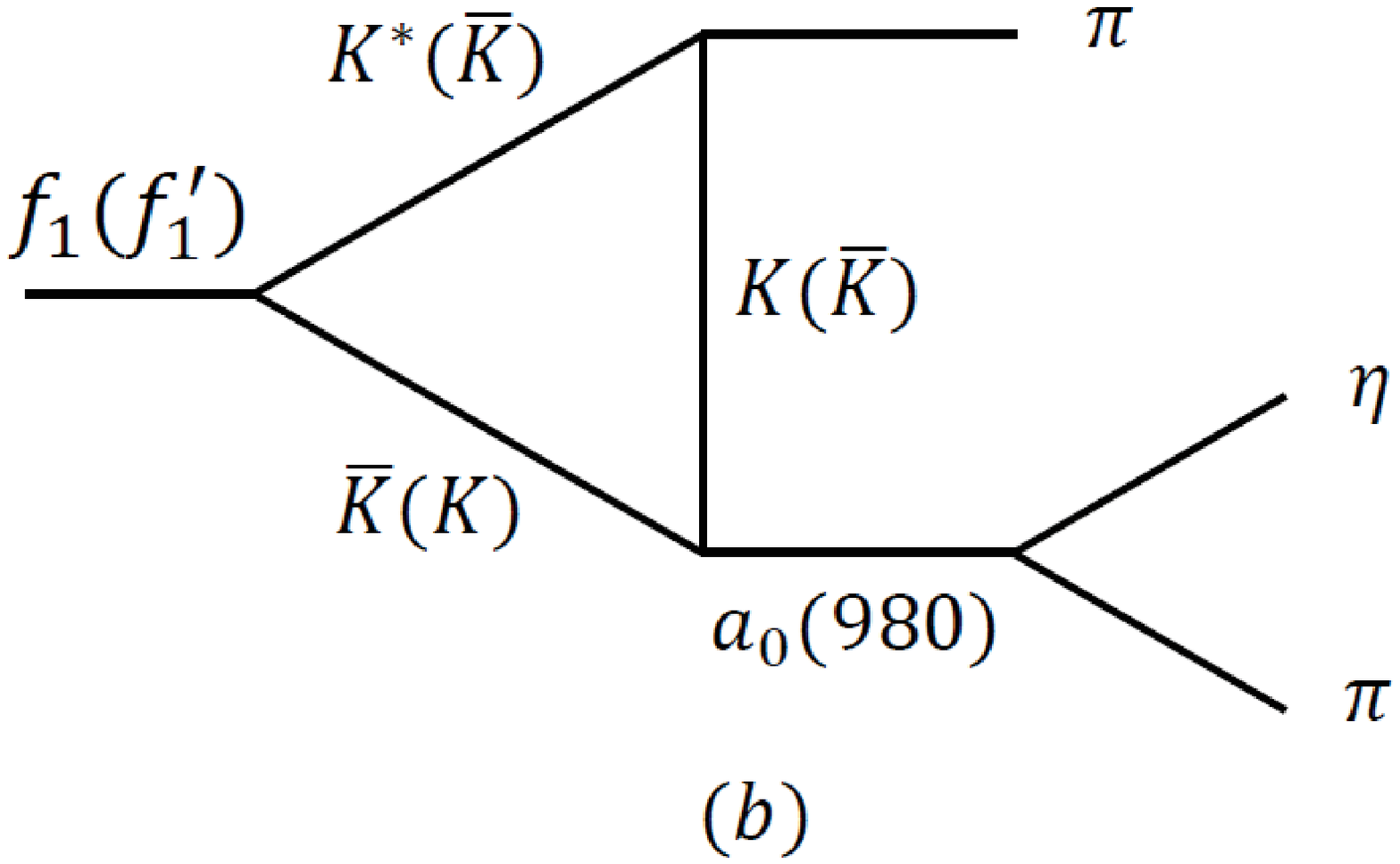}
  \caption{Diagrams for $f_1(f_1')\to a_0\pi\to\eta\pi\pi$.}\label{diagramf1toetapipi}
\end{figure}

\begin{figure}
  \centering
  \includegraphics[width=2in]{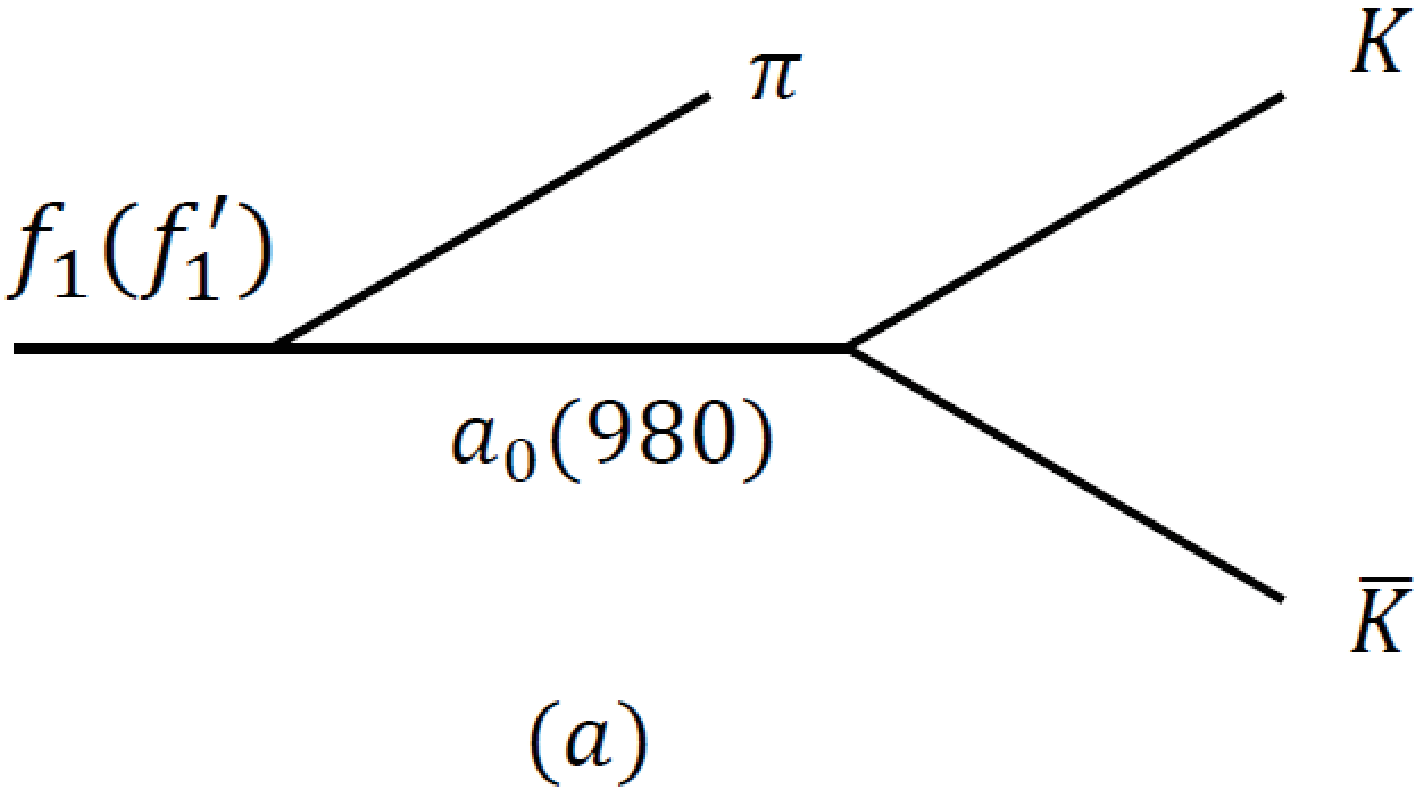}
  \includegraphics[width=2in]{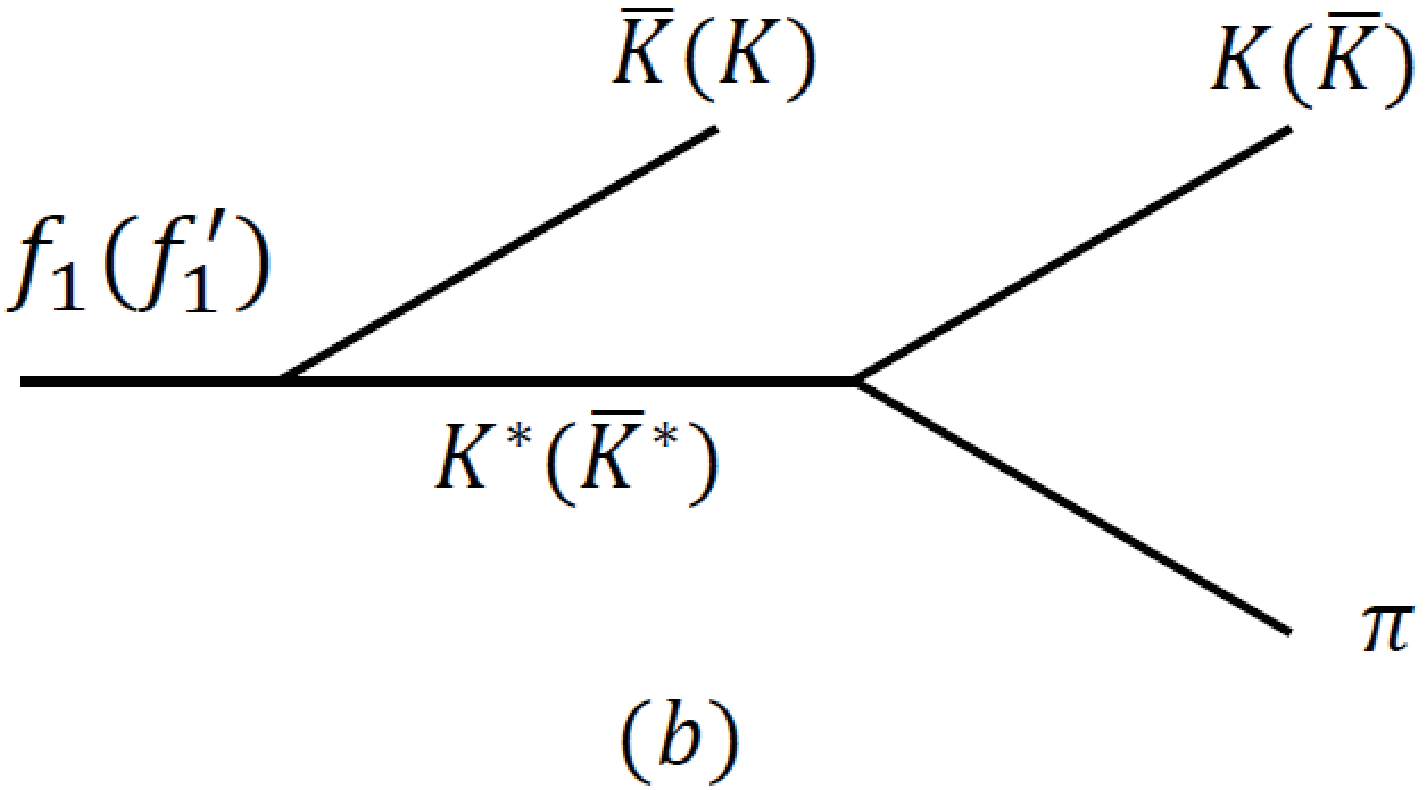}
  \includegraphics[width=2in]{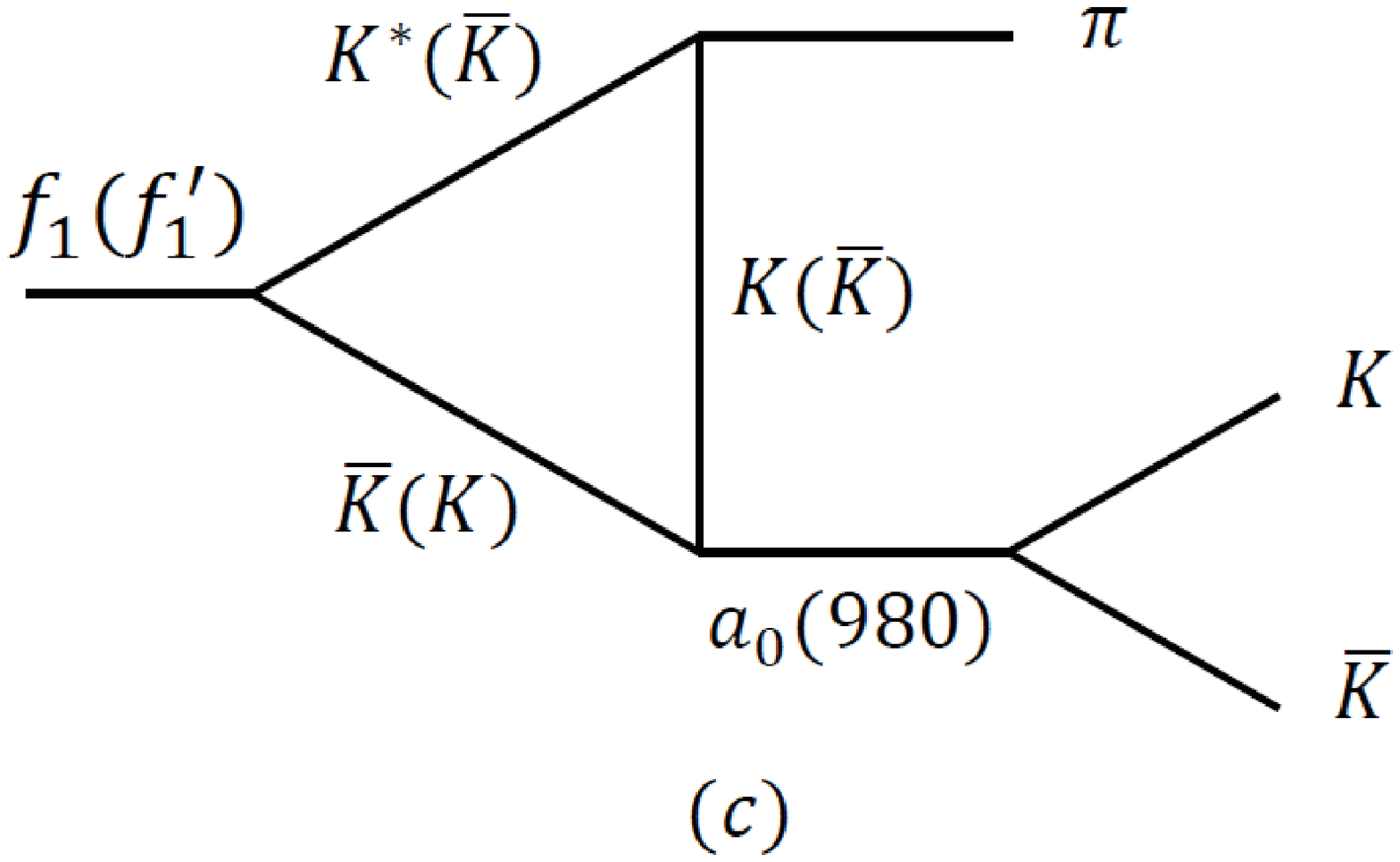}
  \caption{Diagrams for $f_1(f_1')\to K\bar{K}\pi$.}\label{diagramf1tokkpi}
\end{figure}

The decay mechanisms for $f_1$ and $f_1'$ into $\eta\pi\pi$, $K\bar{K}\pi$, and $3\pi$ are grouped into two types. One is the tree-level transitions and the other one is loop transitions via triangle rescatterings. We first illustrated the mechanisms by Feynman diagrams in Figs.~\ref{diagramf1toetapipi} and \ref{diagramf1tokkpi} for the decay channels into $\eta\pi\pi$ and $K\bar{K}\pi$. The isospin-violating decay of $f_1 / f_1'\to 3\pi$ is illustrated by Fig.~\ref{diagramf13pi} and will be discussed later.

For both decay channels of $\eta\pi\pi$ and $K\bar{K}\pi$ the tree-level couplings are originated from the quark model axial vector meson couplings to $a_0(980)\pi$ in a $P$-wave and to $K^*\bar{K}+c.c.$ in an $S$-wave. The strong $S$-wave $K^*\bar{K}$ interaction will have two major physical consequences. One is to dress the bare states and the masses of the physical states correspond to the experimentally measured values. The second one is that the physical couplings for $f_1$ ($f_1'$) to $a_0\pi$ will be renormalized by the triangle transitions. In the $f_1'$ decays the kinematics of the triangle transitions are located within the physical range of the TS mechanism. Thus, one realizes that the inclusion of the TS contributions is necessary. It implies that the physical couplings defined for $f_1 \ (f_1')\to a_0(980)\pi$ should include both tree-level couplings and loop contributions. They will be constrained by experimental data in the combined analysis.

%It should also be stressed that the data from $J/\psi\to \gamma K^\pm K_s^0\pi^\mp$ suggest that the signals for $a_0(980)$ in the invariant mass spectrum of $K\bar{K}$ is very weak~\cite{Bai:2000ss}. In contrast, the signal for $f_1$ in the $a_(980)\pi$ channel is clear. It means that the interference between the tree-level and triangle amplitudes can be possibly constrained.

The following typical Lagrangians are needed in the transition amplitudes for Figs.~\ref{diagramf1toetapipi} and \ref{diagramf1tokkpi}:

(i) Following the convention of Eq.~(\ref{mixingangleoff1}) the Lagrangian for the $f_1 \ (f_1')$ coupling to the vector ($V$) and pseudoscalar ($P$) mesons is expressed as
\begin{eqnarray}
L_1=ig_{AVP}Tr[A^{\mu}[V_{\mu},P]].\label{lagrangianAVP}
\end{eqnarray}
where $A$ denotes the axial vector matrix on the SU(3) basis, i.e.
%with $J^{PC}=1^{++}$ for the neutral nonstrange states
\begin{eqnarray}
A^{\mu}=\left(
\begin{array}{ccc}
 \frac{\cos{\alpha_f}f_1'+\sin{\alpha_f}f_1}{\sqrt{2}}+\frac{a_1^0}{\sqrt{2}} & a_1^+ & K_{1A}^+\\
  a_1^- & \frac{\cos{\alpha_f}f_1'+\sin{\alpha_f}f_1}{\sqrt{2}}-\frac{a_1^0}{\sqrt{2}} & K_{1A}^0 \\
  K_{1A}^- & \bar{K}_{1A}^0 & -\sin{\alpha_f}f_1'+\cos{\alpha_f}f_1
\end{array}
\right).
\end{eqnarray}
The resulting Lagrangians for $f_1(f_1')\to K^*\bar{K}$ are
\begin{eqnarray}
L_{f_1}=ig_{AVP}\frac{-2\cos{\alpha_f}+\sqrt{2}\sin{\alpha_f}}{2}f_1^{\mu}(K^{*0}_{\mu}\bar{K}^0-\bar{K}^{*0}_{\mu}K^0+K^{*+}_{\mu}K^- -K^{*-}_{\mu}K^+),\\
L_{f_1'}=ig_{AVP}\frac{\sqrt{2}\cos{\alpha_f}+2\sin{\alpha_f}}{2}f_1'^{\mu}(K^{*0}_{\mu}\bar{K}^0-\bar{K}^{*0}_{\mu}K^0+K^{*+}_{\mu}K^- -K^{*-}_{\mu}K^+) \ .
\end{eqnarray}
This allows us to define the leading order physical couplings for $f_1 / f_1'$ to $K^*\bar{K}$, i.e.
\begin{eqnarray}\label{coupling-ftoKKstar}
g_{f_1K^{*0}\bar{K}^0}&\equiv & ig_{AVP}\frac{-2\cos{\alpha_f}+\sqrt{2}\sin{\alpha_f}}{2} \ ,\nonumber\\
g_{f_1'K^{*0}\bar{K}^0}&\equiv & ig_{AVP}\frac{\sqrt{2}\cos{\alpha_f}+2\sin{\alpha_f}}{2}.
\end{eqnarray}
With $\theta_{K_1}=34^{\circ}$ adopted as the input, the mixing angle has a value of $\alpha_f=90^{\circ}-5.52^{\circ}=84.48^\circ$. 

(ii) The Lagrangians for $f_1 \ (f_1')$ coupling to $a_0\pi$ have the following form: 
\begin{eqnarray}
L_{f_1a_0\pi} &=&g_{ASP}\sin{\alpha_f}f_1^{\mu}(\pi\partial_{\mu}a_0-a_0\partial_{\mu}\pi)\ ,\\
L_{f_1'a_0\pi} &=&g_{ASP}\cos{\alpha_f}f_1'^{\mu}(\pi\partial_{\mu}a_0-a_0\partial_{\mu}\pi) \ .
\end{eqnarray}
With $\alpha_f=84.48^\circ$ the coupling $g_{f_1'a_0\pi}$ is significantly suppressed with respect of $g_{f_1a_0\pi}$ by a factor of $\cot{\alpha_f}\simeq 0.1$. In the numerical calculations we find that Fig.~\ref{diagramf1tokkpi} (a) can be neglected in $f_1'\to K\bar{K}\pi$. 
The TS effect and the intermediate $a_0(980)$ resonance is able to influence the $K\bar{K}$ spectrum in $f_1'\to K\bar{K}\pi$ at the lower end. But the significance would depend on the relative phase between the two remained amplitudes (i.e. Fig.~\ref{diagramf1tokkpi} (b) and (c)), which arises from the coupling $g_{a_0K\bar{K}}$. This phase can be obtained in $J/\psi\to \gamma\eta(1405)\to \gamma K\bar{K}\pi$ as shown in Ref.~\cite{Du:2019idk}. 

As follows, we provide the transition amplitudes for each decay processes in Figs.~\ref{diagramf1toetapipi} and \ref{diagramf1tokkpi}.

\subsubsection{$f_1'/f_1\to\eta\pi\pi$}  

In this subsection we discuss the detailed transition mechanisms for $f_1'\to \eta\pi\pi$. Then, the corresponding formulas for $f_1\to \eta\pi\pi$ can be obtained by a simple replacement of the vertex coupling constants.

For $f_1'\to \eta\pi\pi$ the amplitudes of the tree (Fig.~\ref{diagramf1toetapipi} (a)) and triangle diagram (Fig.~\ref{diagramf1toetapipi} (b)) have the following expressions, respectively,
\begin{eqnarray}
M_{\eta\pi\pi}^{tree}&=&\epsilon_{f_1'\mu}\frac{1}{\sqrt{2}}ig_{f'a_0\pi}ig_{a_0\eta\pi^0}\frac{i}{D_{a_0}}i(p_a^{\mu}+p_d^{\mu}-p_b^{\mu})+(b\leftrightarrow d).
\end{eqnarray}
and
\begin{eqnarray}\label{loop-amp-etapipi}
M_{\eta\pi\pi}^{loop}&=&\epsilon_{f_1'\mu}\frac{1}{\sqrt{2}}ig_{f_1'K^{*0}\bar{K}^0}ig_{K^{*0}K^0\pi^0}ig_{a_0K^0\bar{K}^0}ig_{a_0\eta\pi^0}2(\hat{I}^{(n)\mu}+\hat{I}^{(c)\mu})\frac{i}{D_{a_0}}+(b\leftrightarrow d) \nonumber\\
&=&\epsilon_{f_1'\mu}\frac{1}{\sqrt{2}}ig_{f_1'K^{*0}\bar{K}^0}ig_{K^{*0}K^0\pi^0}ig_{a_0K^0\bar{K}^0}ig_{a_0\eta\pi^0}4 \hat{I}^{\mu}\frac{i}{D_{a_0}}+(b\leftrightarrow d),
\end{eqnarray}
where $\hat{I}^{(n)\mu}$ and $\hat{I}^{(c)\mu}$ are the loop functions defined for the triangle diagrams in Figs.~\ref{diagramf1toetapipi} and \ref{diagramf1tokkpi}. The superscripts $n$ and $c$ denote the charge-neutral or charged  intermediate mesons. We will give the expression of $\hat{I}^{\mu}$ later in this section. The difference between $\hat{I}^{(n)\mu}$ and $\hat{I}^{(c)\mu}$ is due to the slightly different masses between the charged and neutral kaons, or between the charged and neutral $K^*$. We will see later that this gives a novel source of isospin breaking via the TS mechanism. In the above equation the $\eta$ momentum is labelled by $p_a$, and the momenta for the rest two pions by $p_b$ and $p_d$, respectively. We also note that for $f'\to\eta\pi\pi$, Fig.~\ref{diagramf1toetapipi} (a) is insignificant compared to the triangle amplitude for $f'\to a_0\pi\to\eta\pi^0\pi^0$ (Fig.~\ref{diagramf1toetapipi} (b)).

The $a_0(980)$ resonance is described by a unitary propagator
\begin{eqnarray}\label{a0-propagator}
\frac{i}{D_{a_0(k^2)}}&\equiv &\frac{i}{k^2-m_{a_0}^2-i \sum_{ab}g_{a_0ab}^2\Pi_{ab(k^2)}},
\end{eqnarray}
where
\begin{eqnarray}
\Pi_{ab(k^2)}&\equiv &\int\frac{d^4q}{(2\pi)^4}\frac{1}{((q-k)^2-m_a^2)(q^2-m_b^2)},
\end{eqnarray}
and $ab\in\{\eta\pi, K^0\bar{K}^0, K^+K^-\}$, $g_{a_0ab}$ is the coupling constant of $a_0(980)\to a+b$ and $k$ is the four-vector momentum of $a_0$. We mention in advance that a similar treatment will also be adopted for $f_0(980)$ in this work, for which the propagator is
\begin{eqnarray}\label{f0-propagator}
\frac{i}{D_{f_0(s)}}=\frac{i}{s-m_{f_0}^2-i\sum_{ab}g_{f_0ab}^2\Pi_{ab}}
\end{eqnarray}
with $ab=\{K^+K^-,K^0\bar{K}^0,\pi^+\pi^-,\pi^0\pi^0\}$.

For the decays of $f_1$, the expression is similar and we just need to replace $g_{f_1'K^{*0}\bar{K}^0}$ and $g_{f_1'a_0\pi}$ by $g_{f_1K^{*0}\bar{K}^0}$ and $g_{f_1a_0\pi}$, respectively.

\subsubsection{$f_1'\to K\bar{K}\pi$}

For $f_1'\to K\bar{K}\pi$ we denote the momenta of $K$, $\bar{K}$ and $\pi$ as $p_a$, $p_d$ and $p_b$, respectively. Thus, the transition amplitudes for the processes of Fig.~\ref{diagramf1tokkpi} can be obtained as follows:
\begin{eqnarray}
M_{K\bar{K}\pi}^{tree-1}&=&\epsilon_{f_1'\mu}ig_{f_1'a_0\pi}ig_{a_0 K^0\bar{K}^0}\frac{i}{D_{a_0}}i(p_a^{\mu}+p_d^{\mu}-p_b^{\mu}),\\
M_{K\bar{K}\pi}^{tree-2}&=&ig_{f_1'K^{*0}\bar{K}^0}ig_{K^{*0}K^0\pi}\epsilon_{f_1'\mu}\frac{i}{D_{K^*}}(-g^{\mu\nu}+\frac{q^{\mu}q^{\nu}}{q^2})i(p_a^{\mu}-p_b^{\mu})+(a\rightarrow d),
\end{eqnarray}
with $q=p_a+p_d$, and
\begin{eqnarray}\label{loop-amp-kkpi}
M_{K\bar{K}\pi}^{loop}&=& \epsilon_{f_1'\mu}ig_{f_1'K^{*0}\bar{K}^0}ig_{K^{*0}K^0\pi^0}ig_{a_0K^0\bar{K}^0}ig_{a_0K^0\bar{K}^0}2(\hat{I}^{(n)\mu}+\hat{I}^{(c)\mu})\frac{i}{D_{a_0}} \nonumber\\
&=& \epsilon_{f_1'\mu}ig_{f_1'K^{*0}\bar{K}^0}ig_{K^{*0}K^0\pi^0}ig_{a_0K^0\bar{K}^0}ig_{a_0K^0\bar{K}^0}4 \hat{I}^{\mu}\frac{i}{D_{a_0}}.
\end{eqnarray}

The total amplitude is
\begin{eqnarray}
M_{f'\to K\bar{K}\pi}=M_{K\bar{K}\pi}^{tree-1}+ M_{K\bar{K}\pi}^{tree-2}+M_{K\bar{K}\pi}^{loop}. \label{mf14kkpisum}
\end{eqnarray}
%With the convention in Eq.~(\ref{mixingangleoff1}), the couplings $f_1(f_1')\to K^*\bar{K}$ are obtained from the Lagrangian
Note that the contribution of $M_{K\bar{K}\pi}^{tree-1}$ is much smaller than the other two terms due to the small $f_1'a_0\pi$ coupling. So, the sum of Fig.~\ref{diagramf1tokkpi} (b) and (c) can be a reasonable approximation for $f'\to K\bar{K}\pi$.

\subsubsection{$f_1'\to f_0(980)\pi\to\pi^+\pi^-\pi^0$}

\begin{figure}
  \centering
  \includegraphics[width=2in]{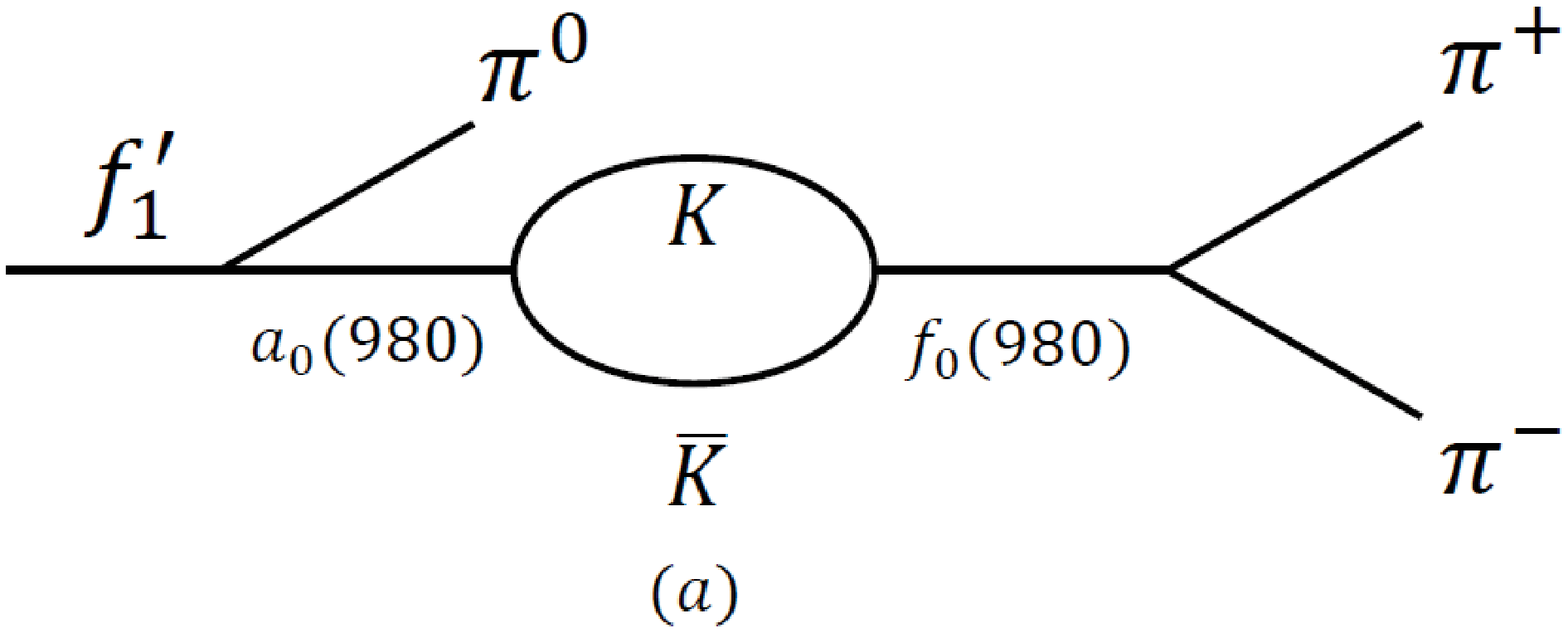}\hspace{1cm}
  \includegraphics[width=2in]{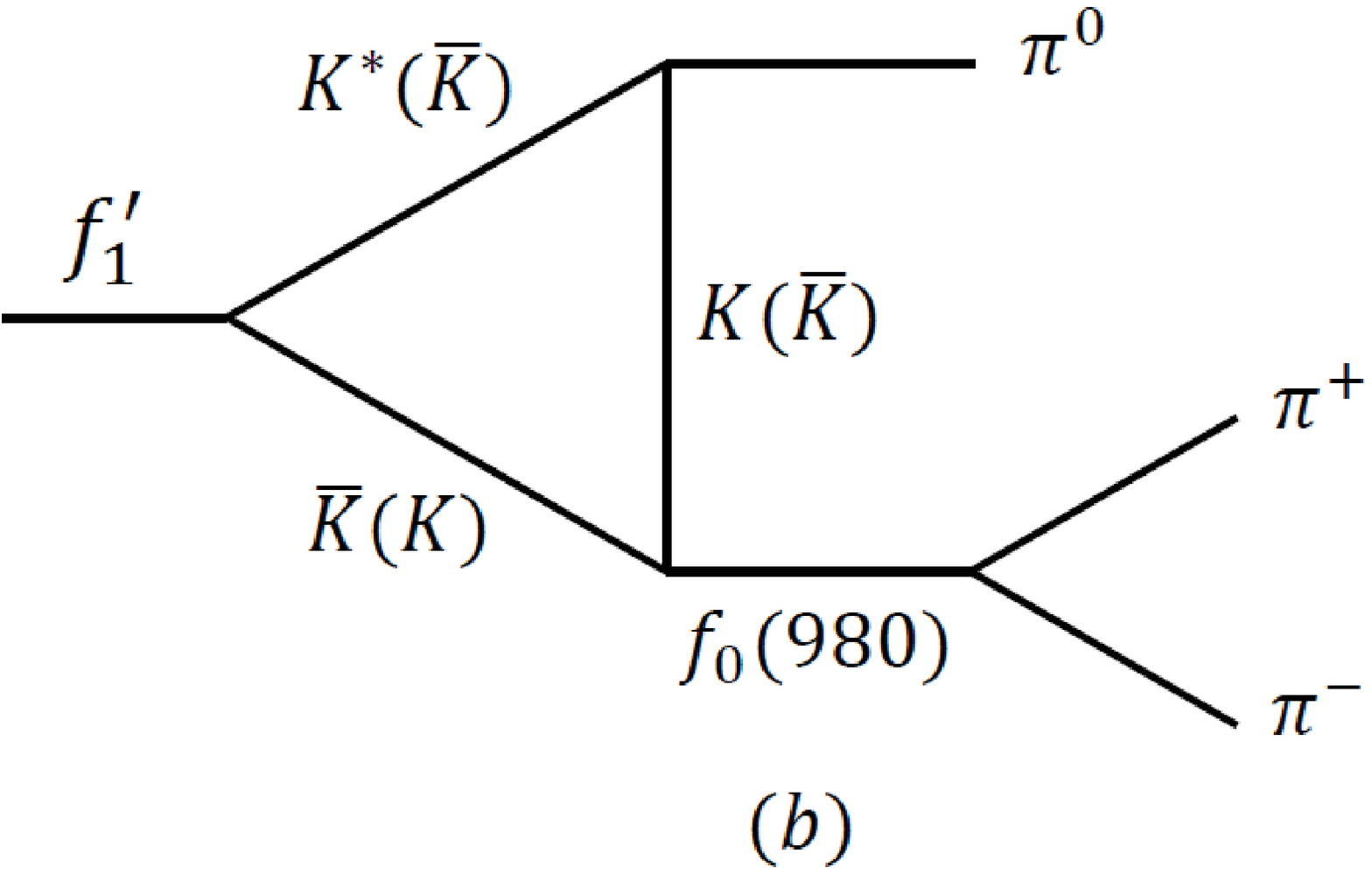}
  \includegraphics[width=2in]{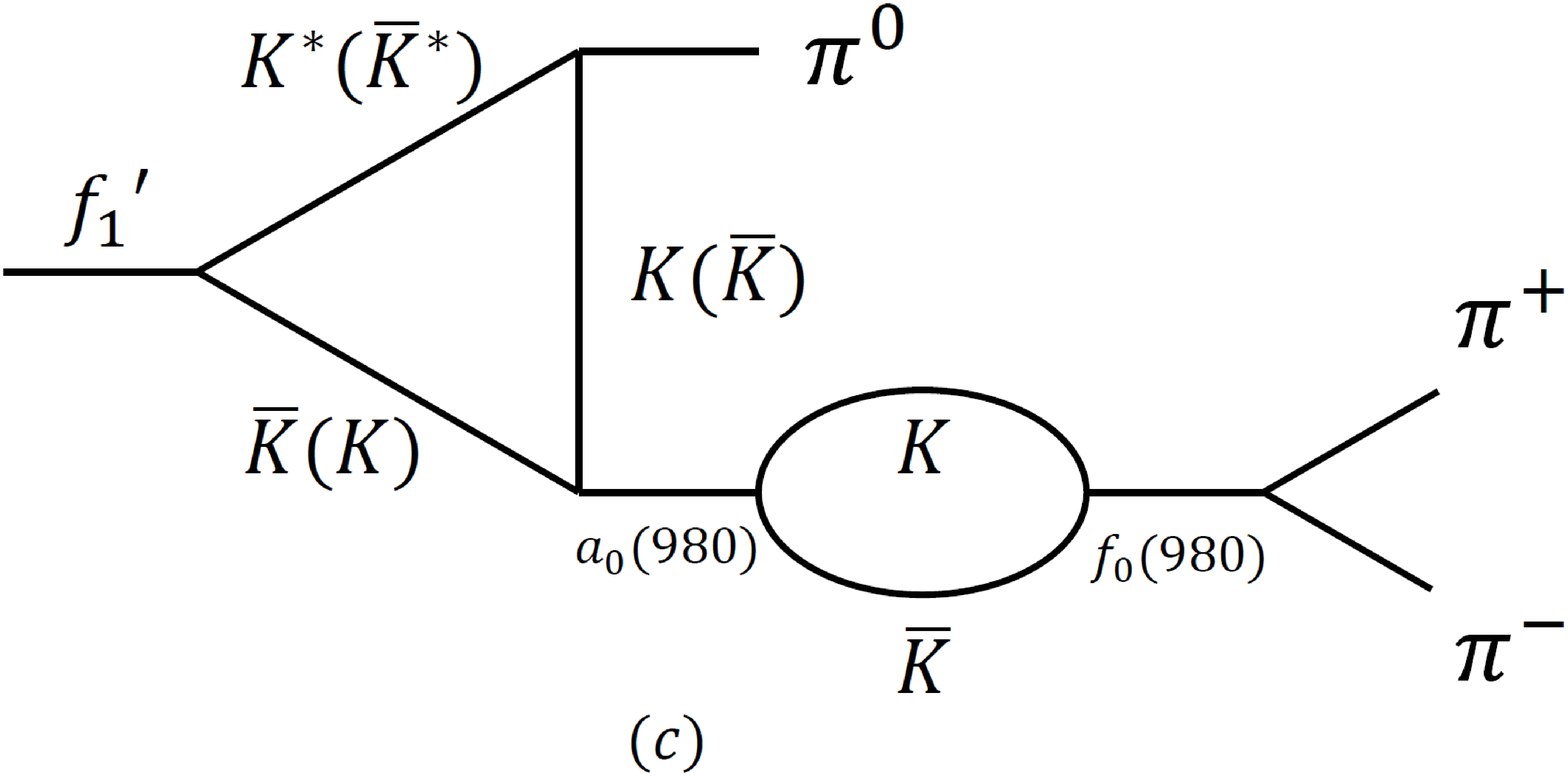}
  \caption{Diagrams for $f_1'\to f_0(980)\pi\to\pi^+\pi^-\pi^0$ channel.}\label{diagramf13pi}
\end{figure}

The isospin-violating process $f_1'\to f_0(980)\pi\to\pi^+\pi^-\pi^0$ can go through the $a_0-f_0$ mixing via a tree-level $a_0$ production (Fig.~\ref{diagramf13pi} (a)) and the triangle mechanism (Figs.~\ref{diagramf13pi} (b) and (c)). This is very similar to the case of the isospin-violating decay of $\eta(1405/1475)\to 3\pi$~\cite{Wu:2011yx,Wu:2012pg,Du:2019idk}. In Ref.~\cite{Wu:2012pg} the process $f_1'\to 3\pi$ in $J/\psi\to\gamma +3\pi$ was first investigated which contains the contributions from Figs.~\ref{diagramf13pi} (a) and (b). It was also found that Fig.~\ref{diagramf13pi} (b) dominates over (a), i.e. the isospin-breaking effects are mainly from the TS mechanism. An improvement of the study of $\eta(1405/1475)\to 3\pi$ by Ref.~\cite{Du:2019idk} suggests that the contributions from Fig.~\ref{diagramf13pi} (c) (the initial state will be $\eta(1405/1475)$ in this case) is also non-negligible. Due to the TS mechanism the production of $a_0$ will be strongly enhanced by the triangle diagram. Thus, the $a_0-f_0$ mixing will also be enhanced. In this work we will include the mechanism of Fig.~\ref{diagramf13pi} (c) as a complete and self-consistent treatment of the isospin-violating decays of $f_1$ and $f_1'$.

The decay amplitudes of Figs.~\ref{diagramf13pi} (a), (b) and (c) read as follows, respectively,
\begin{equation}
M_{f_1\to 3\pi}^{tree mix}=ig_{f_1a_0\pi}ig_{a_0K^+K^-}ig_{f_0\pi^+\pi^-}\frac{i}{D_{a_0}}(\hat{\mathcal{I}}^{(c)}-\hat{\mathcal{I}}^{(n)})\frac{i}{D_{f_0}} \ ,\label{ampf1-tree-mixing}
\end{equation}
and
\begin{eqnarray}
M_{f_1\to 3\pi}^{loop}=2ig_{f_1K^{*0}\bar{K}^0}ig_{K^{*0}K^0\pi^0}ig_{f_0K^0\bar{K}^0}ig_{f_0\pi^+\pi^-}\epsilon_{f\mu}(\hat{I}^{(n)\mu}-\hat{I}^{(c)\mu})\frac{i}{D_{f_0}},\label{ampf1tof0pito3pi}
\end{eqnarray}
and 
\begin{equation}
M_{f_1\to 3\pi}^{loop+mix}=2ig_{f_1K^{*0}\bar{K}^0}ig_{K^{*0}K^0\pi^0}(ig_{a_0K^0\bar{K}^0})^2ig_{f_0K^0\bar{K}^0}ig_{f_0\pi^+\pi^-}\epsilon_{f\mu}(\hat{I}^{(n)\mu}+\hat{I}^{(c)\mu})\frac{i}{D_{a_0}}(\hat{\mathcal{I}}^{(n)}-\hat{\mathcal{I}}^{(c)})\frac{i}{D_{f_0}}.\label{ampf1to3pi-TS-mixing}
\end{equation}
In Eqs.~(\ref{ampf1-tree-mixing}) and ~(\ref{ampf1to3pi-TS-mixing}) functions $\hat{\mathcal{I}}^{(n)}$ and $\hat{\mathcal{I}}^{(c)}$ represent the two-point loop functions with the neutral $K^0\bar{K}^0$ and charged $K^+K^-$ as intermediate particles, respectively, for the $a_0-f_0$ mixing, i.e.
\begin{equation}
\hat{\mathcal{I}}^{(n/c)}=\int\frac{d^4q}{(2\pi)^4}\frac{i^2}{(q^2-m_K)[(p-q)^2-m_K^2]}\ ,
\end{equation}
where the superscripts ``n" and ``c" denote the neutral and charged kaon pairs in the loop function. One can see that these two loop functions cancel each other due to a sign difference between the products of $g_{a_0K^0\bar{K}^0}g_{f_0K^0\bar{K}^0}$ and $g_{a_0K^+\bar{K}^-}g_{f_0K^+\bar{K}^-}$ in the SU(3) flavor symmetry. Thus, the isospin breaking effects are given by the small non-vanishing part caused by the mass difference between the charged and neutral kaon pairs. The propagators of $a_0$ and $f_0$ have been given in Eqs.~(\ref{a0-propagator}) and (\ref{f0-propagator}), respectively.

One qualitative feature of the axial vector productions in the $J/\psi$ radiative decays is that the preferred mixing angle $\alpha_f\sim 90^{\circ}$ implies that Fig.~\ref{diagramf13pi} (a) should be insignificant in comparison with Fig.~\ref{diagramf13pi} (b), and the latter should be the main contribution to the isospin violation in $J/\psi\to \gamma f_1'\to \gamma +3\pi$.

\subsection{$a_1(1260)$ and $a_1(1420)$ decays into $3\pi$}\label{sect:a1}

The strong couplings of the axial vector mesons $f_1$ and $f_1'$ to $K^*\bar{K}+c.c.$ and the presence of the TS mechanism in this kinematic region imply that the $S$-wave isovector coupling between $K^*\bar{K}+c.c.$ can produce observable effects in the $3\pi$ spectrum. 
Experimental evidence is provided by the observation of $a_1(1420)$ at COMPASS in $\pi^- p \to a_1(1420)^\pm\pi^\mp n\to \pi^+\pi^-\pi^0 n$~\cite{Adolph:2015pws}. As a natural consequence of the TS mechanism, it can be accounted for by the strong coupling of the nearby $a_1(1260)$ to the $K^*\bar{K}+c.c.$ threshold~\cite{Ketzer:2015tqa,Aceti:2016yeb}. In the invariant mass spectrum of $3\pi$ the non-vanishing coupling would be strong enough to create the peak of $a_1(1420)$ in the TS kinematic region.

A crucial issue about the role played by the TS mechanism is its relative strength to the $a_1(1260)$ signal in the exclusive decay channel, i.e. $a_1\to 3\pi$. In the picture of TS mechanism for $a_1(1420)$ enhancement, the off-shell coupling of $a_1(1260)$ to the intermediate $K^*\bar{K}+c.c.$ via the triangle loops is actually constrained. Therefore, a combined analysis of $a_1(1260)\to f_0\pi^-\to\pi^+\pi^-\pi^0$ with the production of $a_1(1420)$ will help clarify the nature of $a_1(1420)$.

\begin{figure}
  \centering
  \includegraphics[width=2in]{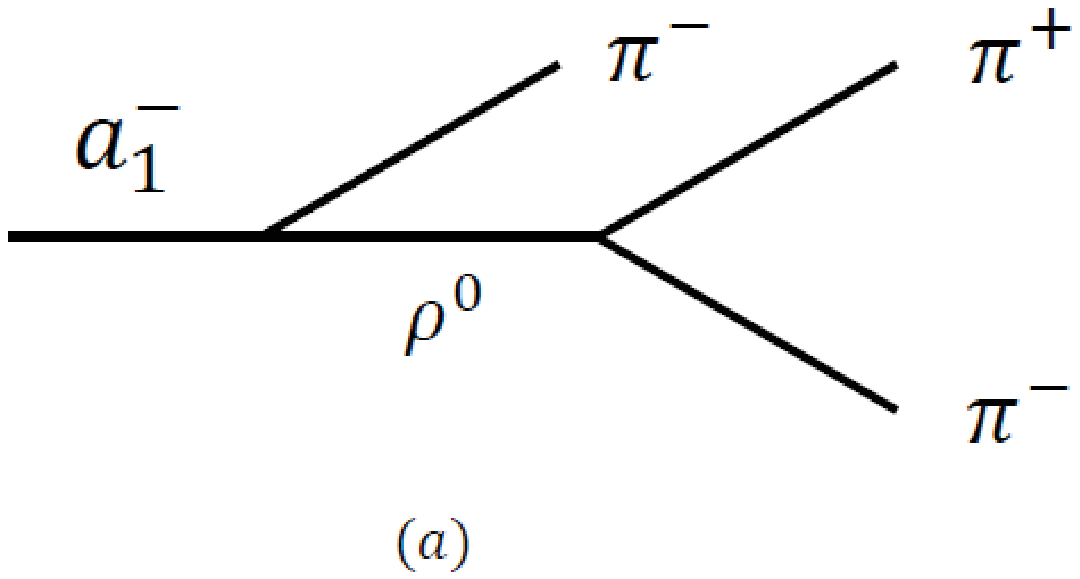}\hspace{1cm}
  \includegraphics[width=2in]{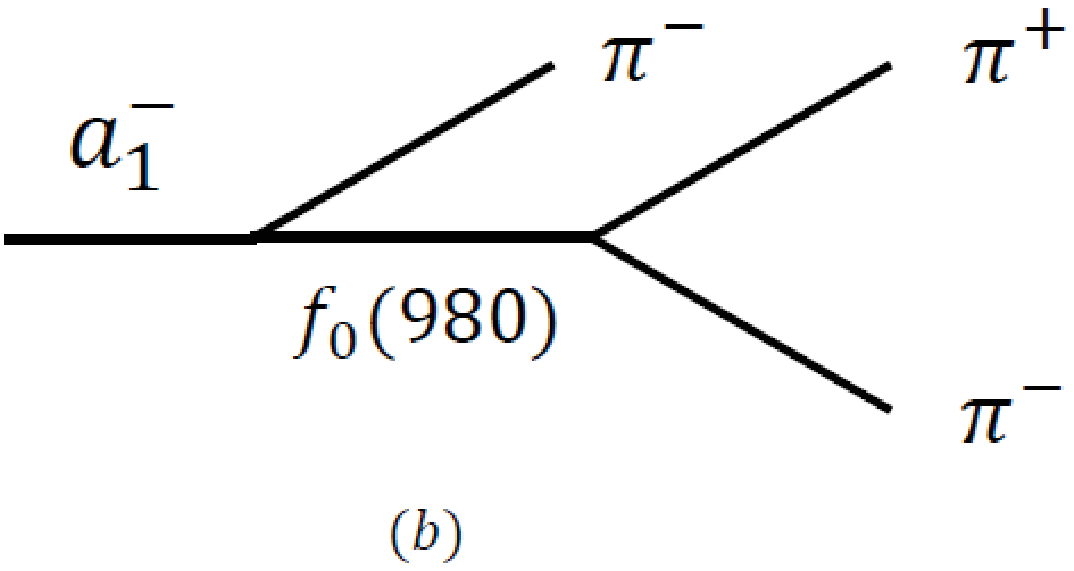}\hspace{1cm}
  \includegraphics[width=2in]{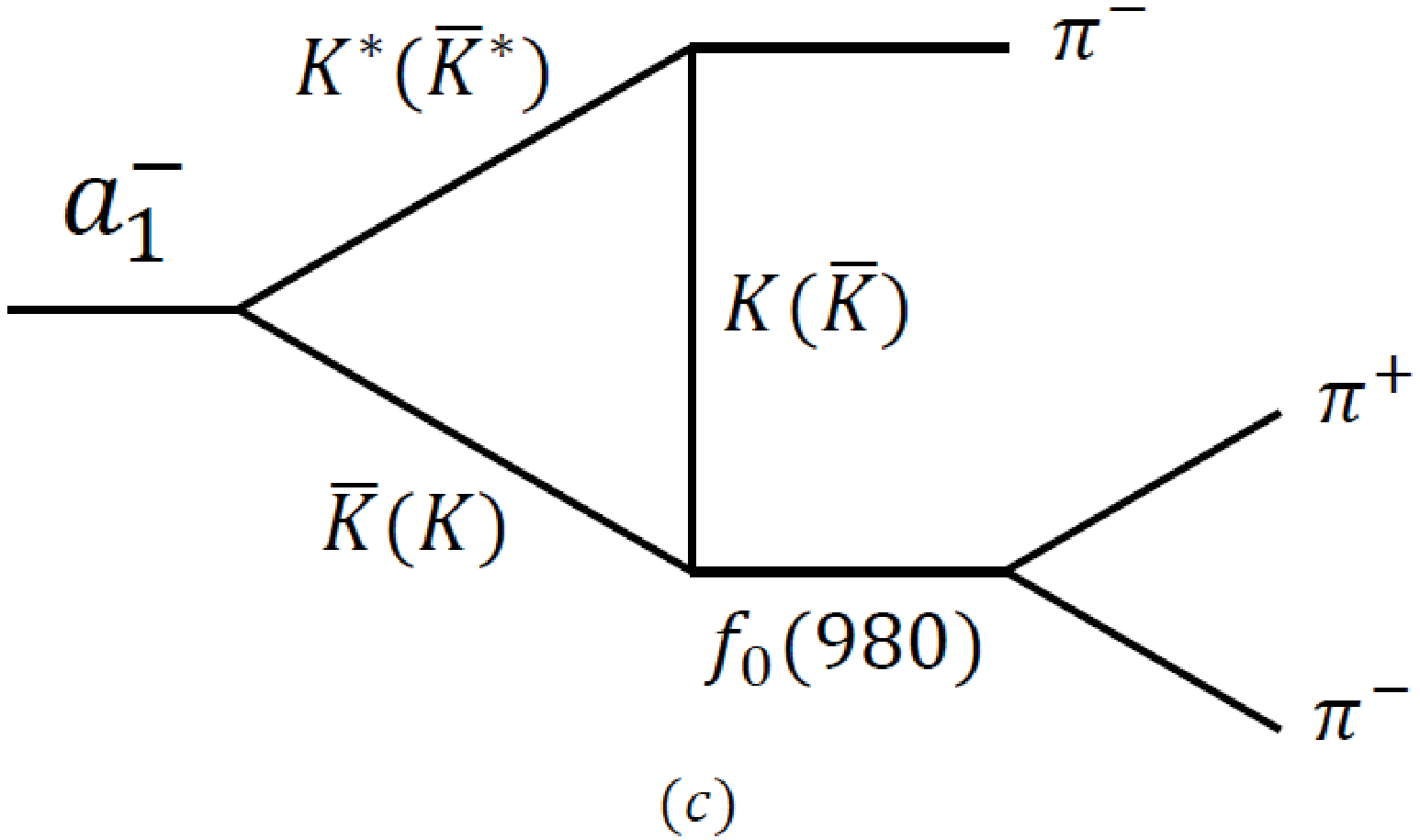}
  \caption{Diagrams for $a_1\to\rho^0\pi^-\to\pi^+\pi^-\pi^-$ (a),$a_1^-\to f_0(980)\pi^-\to\pi^+\pi^-\pi^-$ via the tree (b) and the triangle loop (c) diagrams.}\label{diagram-a1-3pi}
\end{figure}

The transition of $a_1\to 3\pi$ is illustrated by Fig.~\ref{diagram-a1-3pi} which includes two tree diagrams and a triangle process. In Fig.~\ref{diagram-a1-3pi} (a) the transition can go through the intermediate $\rho\pi$ and the corresponding amplitude is
\begin{equation}
M_{a_1\to \rho\pi\to 3\pi}^{tree}=\frac{1}{\sqrt{2}}ig_{a_1^-\rho^0\pi^-}ig_{\rho^0\pi^+\pi^-}\epsilon_{a_1}^{\mu}\frac{i(-g_{\mu\nu}+\frac{p_{c\mu}p_{c\nu}}{p_c^2})}{D_{\rho}}i(p_a^{\nu}-p_{d}^{\nu})+(b\leftrightarrow d),
\end{equation}
where $p_a$ and $p_d$ denote the momenta of $\pi^+$ and $\pi^-$ emitted from the $\rho$ meson with momentum $p_c=p_a+p_d$, respectively. The coupling $g_{a_1^-\rho^0\pi^-}$ can be related to $g_{f_1'K^{*0}\bar{K}^0}$ in the SU(3) flavor symmetry via Eq.~(\ref{lagrangianAVP}). This term plays a role as a background   which can be separated out by a partial wave analysis. Moreover, since it is an $S$-wave-dominant decay, it will not interfere with the transition via the $f_0\pi$ channel. We will see later that the intermediate $\rho\pi$ behaves differently from the $f_0\pi$.

Figure~\ref{diagram-a1-3pi} (b) describes the tree-level transition via the intermediate $f_0\pi$ and the Lagrangian reads,
\begin{equation}
L_{a_1^-f_0\pi^-}=g_{ASP}\sin{\alpha_{f_0}}a_1^{-\mu}(\pi^+\partial_{\mu}f_0-f_0\partial_{\mu}\pi^+),
\end{equation}
The bare coupling between $a_1^-$ and $f_0\pi^-$ is assumed to be proportional to the $n\bar{n}$ component of the $f_0$ meson, which can be described by the mixing angle $\alpha_{f_0}$ between $n\bar{n}$ and $s\bar{s}$ in the scalar meson sector. In this case, the bare coupling strength can be related to the bare coupling between $f_1$ and $a_0\pi$, i.e.$g_{f_1a_0\pi}$. Therefore, the relative phase between the tree and the triangle amplitude of $f_0\pi$ channel can be fixed. The corresponding amplitude is
\begin{equation}\label{amp-tree-a1to3pi}
M_{a_1\to f_0\pi\to 3\pi}^{tree}=\frac{1}{\sqrt{2}}ig_{a_1^-f_0\pi^-}\frac{i}{D_{f_0}}ig_{f_0\pi^+\pi^-}i\epsilon_{a_1}\cdot(p_a+p_d-p_b)+(b\leftrightarrow d) \ .
\end{equation}

Figure~\ref{diagram-a1-3pi} (c) describes the transition via the triangle loop and the amplitudes has the following form,
\begin{eqnarray}
M_{a_1\to 3\pi}^{loop}&=& 2ig_{a_1K^{*0}\bar{K}^0}ig_{K^{*0}K^0\pi^0}ig_{f_0K^0\bar{K}^0}ig_{f_0\pi^+\pi^-}\epsilon_{a\mu}(\hat{I}^{(n)\mu}+\hat{I}^{(c)\mu})\frac{i}{D_{f_0}}+(b\leftrightarrow d). \label{ampa1tof0pito3pi}
\end{eqnarray}
In Eqs.~(\ref{amp-tree-a1to3pi}) and (\ref{ampa1tof0pito3pi}) the momenta of $\pi^+$ and $\pi^-$ emitted from $f_0$ are denoted by $p_a$ and $p_d$, and momentum of the $\pi^-$ which recoils against $f_0$ is $p_b$. In Eq.~(\ref{ampa1tof0pito3pi}) the neutral and charged loop amplitudes are in a constructive phase to be compared with the loop amplitude of Eq.~(\ref{ampf1tof0pito3pi}) for the isospin-violating decay of $f_1\to f_0\pi\to 3\pi$. Note that the combined effect from Eqs.~(\ref{amp-tree-a1to3pi}) and (\ref{ampa1tof0pito3pi}) is to dress up the bare $a_1f_0\pi$ coupling with the triangle loop transition. Similar to $f_1\to a_0\pi$ the sum of these two amplitudes will define the physical coupling which can be extracted from the experimental data for $a_1\to f_0\pi$.

\subsection{Triangle loop function in the axial vector meson decays with $J^{PC}=1^{++}$ }

\begin{figure}
  \centering
  \includegraphics[width=3in]{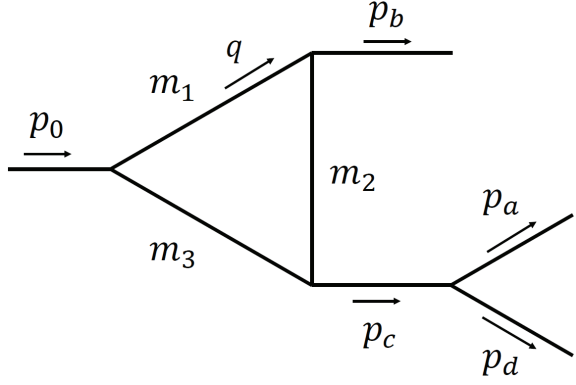}\\
  \caption{Conventions for momenta assignments.}\label{diagramsample}
\end{figure}

In Figs.~\ref{diagramf1toetapipi}, \ref{diagramf1tokkpi}, \ref{diagramf13pi} and \ref{diagram-a1-3pi} the triangle transitions are through the same rescattering processes. The kinematic conditions for the TS mechanism have been discussed in detail in Refs.~\cite{Wu:2011yx,Wu:2012pg,Liu:2015taa,Du:2019idk}. But as pointed out in Ref.~\cite{Du:2019idk}, for different types of vertex couplings in the triangle transitions the TS amplitudes have different structures that leads to different TS effects in these processes. This feature can be seen again for the axial vector decays here. We first consider the triangle amplitude for the axial vector mesons with $C=+1$. The kinematic variables are defined in Fig.~\ref{diagramsample}. The loop function is defined as
\begin{eqnarray}
\hat{I}^{\mu}=\int\frac{d^4q}{(2\pi)^4}\frac{(-g^{\mu\nu}+\frac{q^{\mu}q^{\nu}}{q^2})(q_\nu-2p_{b\nu})F{(q^2)}}{(q^2-m_1^2+i m_1\Gamma_1)((q-p_b)^2-m_2^2)((p_0-q)^2-m_3^2)},\label{loopint1pp}
\end{eqnarray}
where $\{m_1,m_2,m_3\}=\{m_{K^*},m_K,m_{\bar{K}}\}$ and the $\Gamma_1=50$ MeV is the width of $K^*$. The function $F(q^2)$ is the form factor, i.e.
\begin{equation}
F(q^2)=\prod_{i=\{1,2,3\}}\frac{\Lambda_i^2-m_i^2}{\Lambda_i^2-p_{i(q)}^2},
\end{equation}
where $p_{i(q)}$ denotes the four-vector momentum of the intermediate meson of mass $m_i$ which can be expressed as a function of the integration variable $q$; The momenta of the external particles are denoted by $p_a$, $p_d$ and $p_b$, respectively. The cut-off energy is defined by $\Lambda_i\equiv m_i+\beta\Lambda_{QCD}$ ($\Lambda_{QCD}=250 \ \text{MeV}$). Though the original integral without the form factor does converge, it happens often that the dispersive part of the loop amplitude is overestimated when the interacting hadrons are treated as fundamental particles. The introduction of the form factor will then cut off the unphysical ultra-violet contributions in the dispersive part of the loop amplitude.

\section{Production and decays of axial vector mesons with $J^{PC}=1^{+-}$}

\subsection{Productions of the $C=-1$ axial vector mesons in charmonium decays}\label{prod-negative-c}

The axial vector mesons with $C=-1$ can be produced in association with a pseudoscalar meson in the $J/\psi$ hadronic decays, e.g. $J/\psi\to b_1^+\pi^-+c.c.$ and $h_1\eta$.  Then the axial vector mesons can decay into a vector plus a pseudoscalar via either a tree process or a triangle loop transition. There are special advantages with the $b_1(1235)$ and $h_1(1170)/h_1(1415)$ (i.e. $h_1/h_1'$) decays into $\phi\pi$. Since $b_1(1235)$ is an isovector it does not contain the strangeness in the constituents. Therefore, its decays into $\phi\pi$ will be suppressed at the tree level. This provides an ideal place for investigating the TS mechanism which would produce unique signals for the TS mechanism. It is also possible that the production of $J/\psi\to K^*\bar{K}\pi+c.c.$ gives access to the quantum number of $(I, \ J^{P(C)})=(1, \ 1^{+(-)})$ directly via the $K^*\bar{K}+c.c.$ scatterings. Such a possibility has been studied recently by Ref.~\cite{Jing:2019cbw}. 
For $J/\psi\to h_1\eta$ and $h_1'\eta$ with $h_1/h_1'$ decays into $\phi\pi$, the decays of $h_1/h_1'$ violate isospin. Similar to the isospin-violating decays of $f_1\to 3\pi$ or $\eta(1405)\to 3\pi$, this channel is ideal for probing the role played by the TS mechanism as the leading isospin-breaking mechanism. 

For the purpose of detecting the TS mechanism the three-body decays of the axial vector mesons is of interest. For instance, it would be interesting to examine the $K\bar{K}$ invariant mass spectrum in $J/\psi\to b_1^+\pi^-+c.c.\to K\bar{K}\pi^+\pi^-$ since it covers the physical region of the TS. Although the intermediate $\phi$ meson will account for the main cross sections for $b_1\to K\bar{K}\pi$, the impact of the TS mechanism may influence the lineshape of the $K\bar{K}$ spectrum even near the $\phi$ peak. In the following subsections we investigate two correlated processes, i.e. $J/\psi\to b_1^+\pi^-+c.c.\to  K\bar{K}\pi^+\pi^-$ and $J/\psi\to h_1'\eta\to K\bar{K}\pi\eta$ in order to disentangle the role played by the TS mechanism.

The interactions between $J/\psi\to b_1\pi$ and $J/\psi\to h_1'\eta$ can be parametrized by the following Lagrangian in the SU(3) symmetry,
\begin{equation}
L_{\psi BP}=g_{\psi BP}\psi^{\mu}Tr[B_{\mu}P],\label{eq_LpsiBP}
\end{equation}
from which the interaction reads
\begin{eqnarray}
L_{\psi b_1\pi}&=&g_{\psi BP}\psi^{\mu}(b_{1\mu}^+\pi^-+b_{1\mu}^-\pi^+),\label{eq_psib1pi}\\
L_{\psi h_1'\eta}&=&g_{\psi BP}\cos{(\alpha_h-\alpha_P)}\psi^{\mu}h_{1\mu}'\eta,\\
L_{\psi h_1'\eta'}&=&-g_{\psi BP}\sin{(\alpha_h-\alpha_P)}\psi^{\mu}h_{1\mu}'\eta' \ , \label{eq_psih1hetap}\\
L_{\psi h_1\eta}&=&g_{\psi BP}\sin{(\alpha_h-\alpha_P)}\psi^{\mu}h_{1\mu}\eta\ ,\\
L_{\psi h_1\eta'}&=&g_{\psi BP}\cos{(\alpha_h-\alpha_P)}\psi^{\mu}h_{1\mu}\eta'\ .\label{eq_psih1letap}
\end{eqnarray}
The $\alpha_P$ is the mixing angle between $I^{G}J^{PC}=0^+0^{-+}$ $n\bar{n}$ and $s\bar{s}$ states as defined in Eq.~(\ref{pseudoscalar-mixing}). The typical value of $\alpha_P$ is within a range of $38^\circ\sim 42^\circ$ and we adopt $\alpha_P= 40^{\circ}$ in this study. With the mixing angle $\alpha_h=91.77^\circ$, the coupling constants in the decay channels of Eqs.~(\ref{eq_psib1pi})-(\ref{eq_psih1letap}) are compatible with each other. For the purpose of identifying the impact from the TS mechanism, we will focus on the production of $h_1'$ recoiling $\eta$ in the $J/\psi$ decays. This will benefit from a relatively larger phase space factor than $J/\psi\to h_1'\eta'$. 

The coupling strength $g_{\psi BP}$ can be fixed by the branching ratio of $J/\psi\to b_1^{\pm}\pi^{\mp}=(3\pm 0.5)\times 10^{-3}$~\cite{Patrignani:2016xqp}, which leads to $g_{\psi BP}=4.35\times 10^{-3}$ GeV. With $g_{\psi BP}$ and Eq.~(\ref{eq_psih1hetap}), one obtains the branching ratio of $J/\psi\to h_1'\eta'$, i.e. $B.R.(J/\psi\to h_1'\eta')=5.9\times 10^{-4}$. This value is approximately 3 times larger than the experimental measurement of the combined branching ratio $Br^{exp}(J/\psi\to h_1'\eta'\to K^*\bar{K}\eta'+c.c.)=5.9\times 10^{-4}$ by BESIII~\cite{Ablikim:2018ctf}. It implies that the decay of $h_1'\to K^*\bar{K}+c.c.$ is about $1/3\sim 1/2$ of the total width of $h_1'$ which is a reasonable expectation.

\subsection{$J/\psi\to b_1^+\pi^-+c.c.\to K\bar{K}\pi^+\pi^-$}\label{sub:b1toKKpi}

\begin{figure}
  \centering
  \includegraphics[width=2.in]{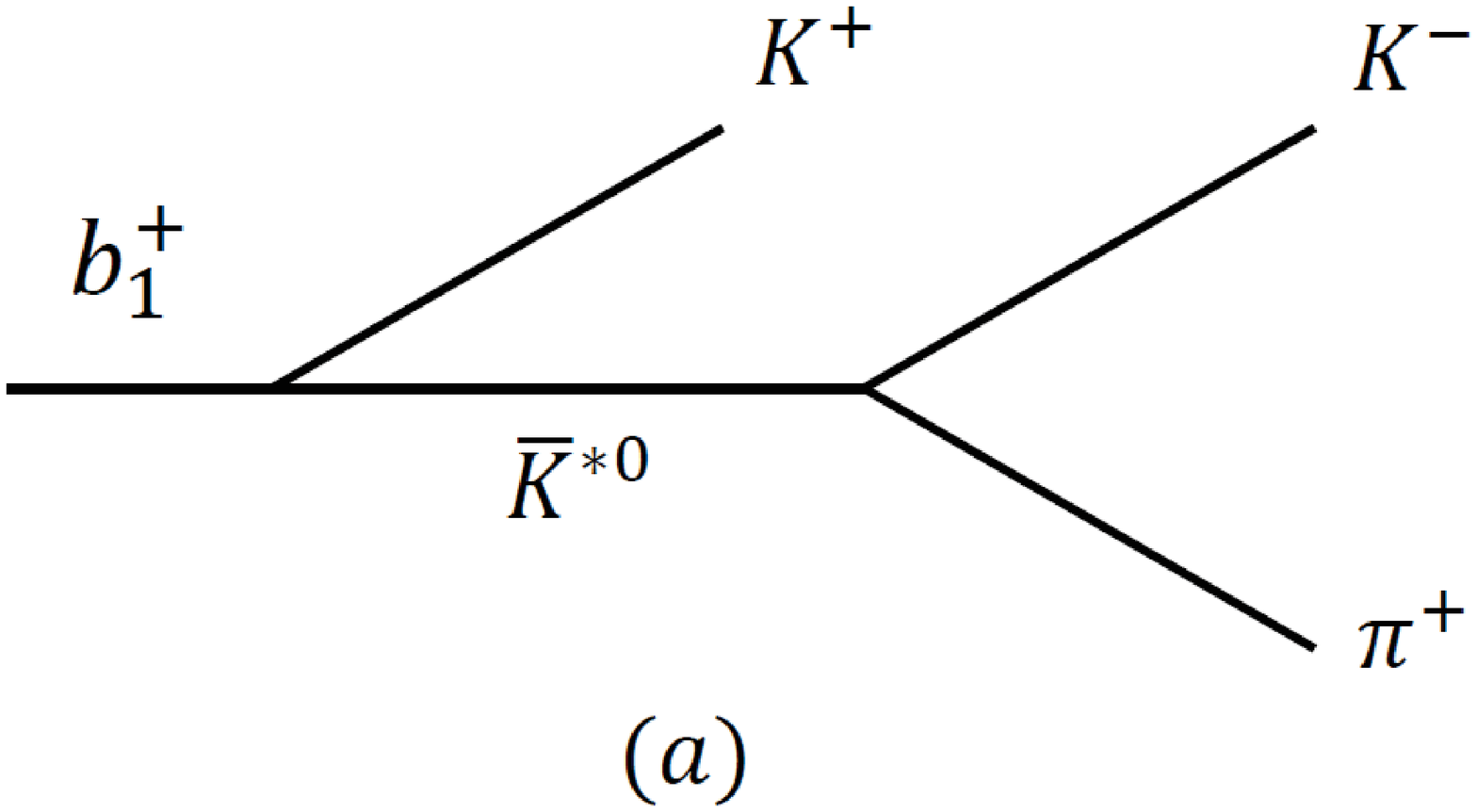}
  \includegraphics[width=2.in]{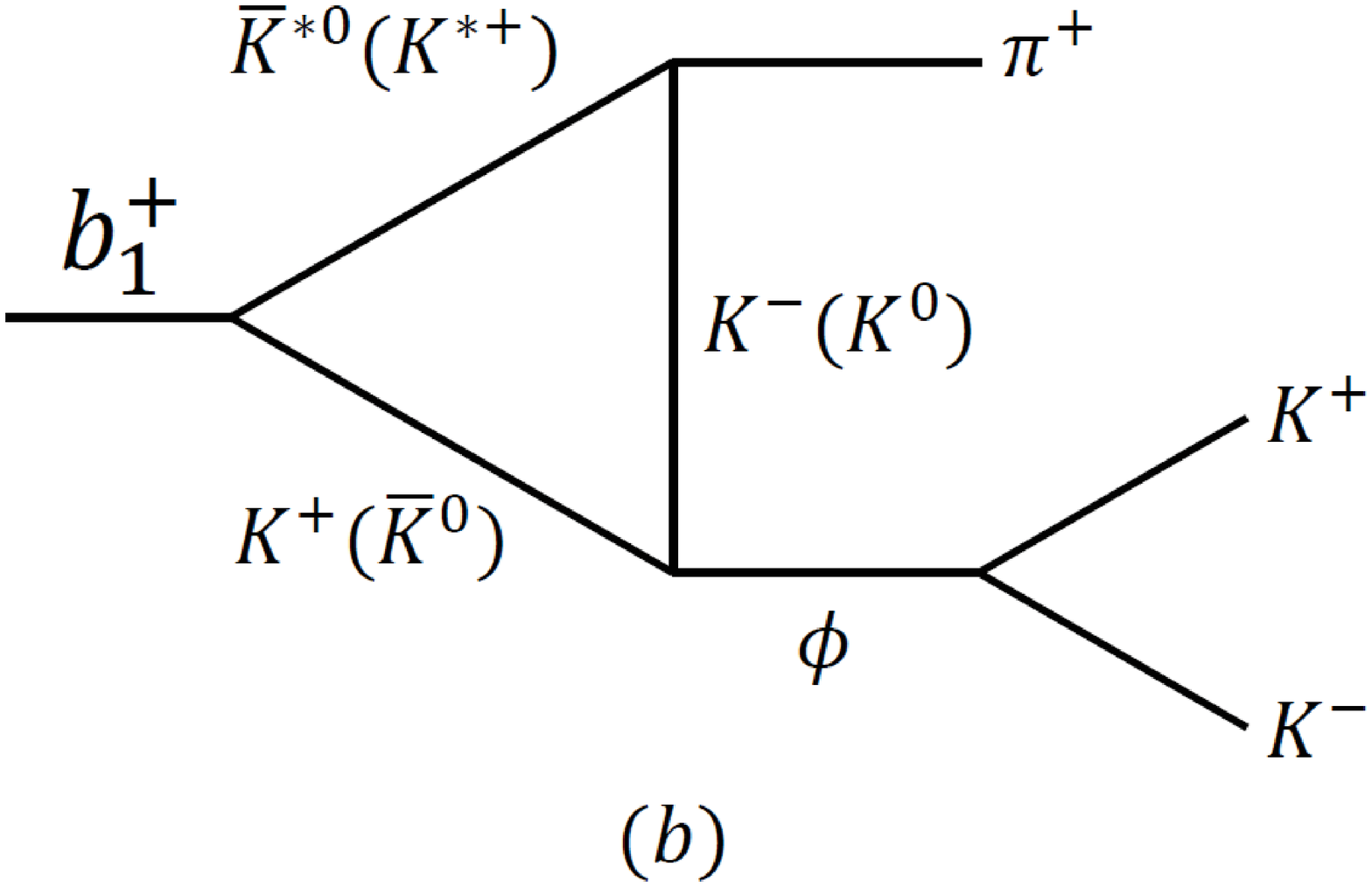}
  \caption{Diagrams of $b_1^+\to K^+K^-\pi^+$ via $K^{*+}\bar{K}^0$ channel (a) and $\phi\pi^+$ channel (b).}\label{diagramb1-to-kkbarpi}
\end{figure}

In $J/\psi\to b_1^+\pi^-+c.c.\to K\bar{K}\pi^+\pi^-$ the $b_1$ decay can go through the processes illustrated in Fig.~\ref{diagramb1-to-kkbarpi}, where (a) shows the tree-level transition via  the intermediate $K^*\bar{K}+c.c.$ and (b) is the loop transition via the intermediate $\phi\pi$. Note that the tree-level decay of $b_1\to\phi\pi$ will be suppressed by the OZI rule. Thus, the triangle loop transition of Fig.~\ref{diagramb1-to-kkbarpi} (b) actually provides a mechanism for evading the OZI rule. Meanwhile, the tree-level transition of Fig.~\ref{diagramb1-to-kkbarpi} (a) behaves as a background contribution to Fig.~\ref{diagramb1-to-kkbarpi} (b). We include this tree-level transition in the analysis of the $b_1^+\to K\bar{K}\pi^+$ in order to have a realistic description of the $K\bar{K}$ invariant mass spectrum with the presence of the TS mechanism.

The amplitude for  $J/\psi\to b_1^+\pi^-+c.c.\to \phi\pi^+\pi^-$ with the processes considered in Fig.~\ref{diagramb1-to-kkbarpi}(b) can then be expressed as 
\begin{equation}
M_{J/\psi\to b_1\pi\to\phi\pi^+\pi^-}=g_{\psi BP}\epsilon_{\psi\mu}\epsilon^*_{\phi\rho}\left[\frac{i(-g^{\mu\nu}+\frac{p_c^{\mu}p_c^{\nu}}{s_c})}{D_{b_1^+}(s_c)}J^{+\rho}_{\nu}+\frac{i(-g^{\mu\nu}+\frac{p_{ab}^{\mu}p_{ab}^{\nu}}{s_{ab}})}{D_{b_1^-}(s_{ab})}J^{-\rho}_{\nu}\right] \ ,
\label{amppsitob1pi}
\end{equation}
where $p_a$ and $p_c$ are the momenta of $\phi$ and $b_1^+$, respectively; $p_b$ is the recoiled $\pi^-$ momentum and $p_{ab}=p_a+p_b$ correspond to the $b_1^-$ momentum in the charge conjugation channel, and $s_c\equiv p_c^2$ and $s_{ab}\equiv p_{ab}^2$; $J^{\pm\rho}_{\nu}$ denotes the current for $b_1^{\pm}\to \phi\pi^{\pm}$; $g_{\psi BP}$ is the coupling constant for $J/\psi\to b_1\pi$ introduced in Eq.~(\ref{eq_psib1pi}). In the above equation the inverse propagator for $b_1$ is
\begin{equation}
D_{b_1}(s)=s-m_{b_1}^2+im_{b_1}\Gamma_{b_1}(s),
\end{equation}
where we have assumed an $S$-wave energy dependence of $\Gamma_{b_1}$, i.e.
\begin{equation}
\Gamma_{b_1}(s)=\frac{m_{b_1}|\vec{p}_{\omega(s)}|}{\sqrt{s}|\vec{p}_{\omega(m_{b_1})}|}\Gamma_{b_1}(m_{b_1}^2), 
\end{equation}
with $\Gamma_{b_1}(m_{b_1}^2)=0.142$ GeV~\cite{Patrignani:2016xqp}.

For $b_1\to K\bar{K}\pi$, the amplitude is parametrized as the following form 
\begin{equation}
M_{b_1^+\to K^+K^-\pi^+}=\epsilon_{b_1\mu}(\lambda_ap_a^{\mu}+\lambda_b p_b^\mu+\lambda_dp_d^{\mu}),
\end{equation}
where $p_a$, $p_d$ and $p_b$ are momenta of $K^+$, $K^-$ and $\pi^+$, respectively. The coefficients $\lambda_i$ contains contributions from Fig.~\ref{diagramb1-to-kkbarpi} (a) and (b).

For the convenience of calculations we provide the amplitudes for $b_1^+\to K^+K^-\pi^+$ as follows for these two processes in Fig.~\ref{diagramb1-to-kkbarpi}. The amplitude of $b_1^+\to K^+\bar{K}^{*0}\to K^+K^-\pi^+$ (Fig.~\ref{diagramb1-to-kkbarpi} (a)) reads
\begin{equation}
M_{b_1^+\to K^+K^-\pi^+}^{tree}=ig_{b_1^+\bar{K}^{*0}K^+}ig_{\bar{K}^0K^-\pi^+}\epsilon_{b_1\mu}\frac{i(-g^{\mu\nu}+\frac{p_{ab}^{\mu}p_{ab}^{\nu}}{s_{ab}})}{D_{\bar{K}^{*0}}}i(p_a-p_b)_{\nu},
\end{equation}
where $p_{ab}=p_a+p_b$ is the momentum of $\bar{K}^{*0}$. 

Similarly, the amplitude of $b_1^+\to\phi\pi^+\to K^+K^-\pi^+$ (Fig.~\ref{diagramb1-to-kkbarpi} (b)) can be obtained
\begin{eqnarray}\label{b1tokkpi-tri}
M_{b_1^+\to K^+K^-\pi^+}^{tri}&=& 2ig_{b_1^+\bar{K}^{*0}K^+}ig_{\bar{K}^{*0}K^-\pi^+}ig_{\phi K^+K^-}ig_{\phi K^+K^-}\epsilon_{b_1\mu}\tilde{I}^{\mu\alpha}\frac{i(-g^{\alpha\beta}+\frac{p_c^{\alpha}p_c^{\beta}}{p_c^2})}{p_c^2-m_{\phi}^2+im_{\phi}\Gamma_{\phi}}i(p_a-p_d)_{\beta}\nonumber\\
&=& ig_{b_1^+\bar{K}^{*0}K^+}ig_{\bar{K}^{*0}K^-\pi^+}ig_{\phi K^+K^-}ig_{\phi K^+K^-}\epsilon_{b_1\mu}(\tilde{I}_n^{\mu\alpha}+\tilde{I}_c^{\mu\alpha})\frac{i(-g^{\alpha\beta}+\frac{p_c^{\alpha}p_c^{\beta}}{p_c^2})}{p_c^2-m_{\phi}^2+im_{\phi}\Gamma_{\phi}}i(p_a-p_d)_{\beta} \ ,
\end{eqnarray}
where the momenta of the $\pi^+$, $\phi$, $K^+$ and $K^-$ mesons are labeled by $p_b$, $p_c$, $p_a$ and $p_d$, respectively; $\tilde{I}_{\mu\alpha}$ is the typical loop function for the triangle loop, i.e.
\begin{eqnarray}
\tilde{I}^{\mu\alpha}\equiv -i\int\frac{d^4q}{(2\pi)^4}\frac{(-g^{\mu\nu}+\frac{q^{\mu}q^{\nu}}{q^2})(q-2p_b)_{\nu}(p_c+2p_b-2q)_{\alpha}F(q^2)}{(q-m_1^2+im_1\Gamma_1)((q-p_b)^2-m_2^2)((p_0-q)^2-m_3^2)},
\label{loopint1pm}
\end{eqnarray}
which is a tensor integral and will be contracted by the polarization vectors of $b_1$ and $\phi$. Considering the slightly different masses between the charged and neutral intermediate states due to the isospin symmetry breaking, we note the charged and neutral loop integral functions by $\tilde{I}_c^{\mu\alpha}$ and $\tilde{I}_n^{\mu\alpha}$, respectively.

In these amplitudes the unknown coupling $g_{b_1^+\bar{K}^{*0}K^+}$ can be calculated by two ways based on the SU(3) symmetry. On the one hand, it can be connected to the decay of $b_1^+\to \omega\pi^+$ where $g_{b_1^+\omega\pi^+}$ can be extracted by assuming that $b_1\to \omega\pi$ exhausts its total width. The partial width for $b_1^+\to\omega\pi^+$ reads
\begin{equation}
\Gamma(b_1^+\to\omega\pi^+)=\frac{|\vec{p}_{\omega}|}{8\pi m_{b_1}^2}\frac{g_{b_1^+\omega\pi^+}^2}{3}(2+\frac{(m_{b_1}^2+m_{\omega}^2-m_{\pi}^2)^2}{4m_{b_1}^2m_{\omega}^2}),
\end{equation}
where the $|\vec{p}_{\omega}|$ refers to the momentum of $\omega$ in the rest frame of $b_1$. 

On the other hand, $g_{b_1^+\bar{K}^{*0}K^+}$ can be calculated by $g_{BVP}$, which is obtained by calculating $h_1\to\rho\pi$. In the SU(3) limit, these two methods should give the same result. The numerical results show that there is about $10\%$ discrepancy between these two methods, which is acceptable. The discrepancy may results from the following aspects:
\begin{itemize}
\item The SU(3) flavour symmetry only approximately holds.
\item The estimation based on $g_{h_1\rho\pi}$ is correlated with the determination of the $h_1$-$h_1'$ mixing. Thus, the mixing angle may introduce uncertainties to the coupling constant.
\item The $b_1\to\omega\pi$ process contains non-negligible $D$-wave contributions, which is not considered in the above estimation.
\end{itemize}
However, we regard that the $10\%$ discrepancy is an indication of self-consistency in the determination of the $h_1$-$h_1'$ mixing angle. So we adopt $g_{BVP}$ extracted from $h_1\to\rho\pi$ in the final numerical calculations.

\subsection{$h_1'\to \phi\pi\to K\bar{K}\pi$}

\begin{figure}
  \centering
  \includegraphics[width=2in]{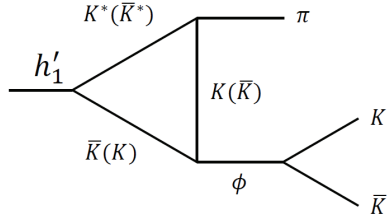}
  \caption{Triangle diagram of $h_1'\to\phi\pi\to K\bar{K}\pi$. The charged and neutral loops are destructive.}\label{diagramh1tophipi}
\end{figure}

The triangle loop transition in Fig.~\ref{diagramh1tophipi} provides an isospin breaking mechanism for $h_1'\to \phi\pi\to K\bar{K}\pi$. The loop integral has the same form as Eq.~(\ref{loopint1pm}), except that the charged and neutral loops must cancel due to the isospin breaking. Similar to the case of the isospin-breaking decay of $f_1'\to 3\pi$, the unequal masses between the intermediate charged and neutral $K$ (or $K^*$) mesons will lead to residue amplitudes after the cancellation. 

The production of $h_1'$ in $J/\psi\to h_1'\eta$ can be described the same way as that for $J/\psi\to b_1\pi$ in the previous subsection. Here, we consider the decay of $h_1'\to \phi\pi$ for which the amplitude can be written as:
\begin{eqnarray}
M_{h_1'\to K^+K^-\pi^0}^{tri}&=& ig_{h_1'K^+\bar{K}^{*-}}ig_{\bar{K}^{*-}K^-\pi^0}ig_{\phi K^+K^-}ig_{\phi K^+K^-}\epsilon_{h_1'\mu}(\tilde{I}_n^{\mu\alpha}-\tilde{I}_c^{\mu\alpha})\frac{i(-g^{\alpha\beta}+\frac{p_c^{\alpha}p_c^{\beta}}{p_c^2})}{p_c^2-m_{\phi}^2+im_{\phi}\Gamma_{\phi}}i(p_a-p_d)_{\beta} \ ,
\end{eqnarray}
which is to be compared with Eq.~(\ref{b1tokkpi-tri}) and the loop function, $\tilde{I}^{\mu\alpha}$ is the same as Eq.~(\ref{loopint1pm}). The inverse propagator for $h_1'$ is
\begin{equation}
D_{h_1'}(s)=s-m_{h_1'}^2+im_{h_1'}\Gamma_{h_1'}(s),
\end{equation}
with
\begin{equation}
\Gamma_{h_1'}(s)\simeq \Gamma_{h_1'\to K^*\bar{K}\to K\bar{K}\pi}(s) \ ,
\end{equation}
where we have assumed that the $K^*\bar{K}$ channel is dominant in the $h_1'$ decays. As shown by Eq.~(\ref{kkpiwidthofh1h}), this is a reasonable approximation concerning the poor experimental status of $h_1$ and $h_1'$.

\section{Results and discussions}

\subsection{Numerical results for the $J^{PC}=1^{++}$ states}

\subsubsection{Parameters}

The parameters in this study include the vertex coupling constants and a cut-off energy for the loop integrals. 
For the positive $C$-parity channel the key parameters are $g_{f_1'K^*\bar{K}}$ and $g_{f_1'a_0\pi}$. They will determine the relative strengths between the $K\bar{K}\pi$ and $\eta\pi\pi$ channel and the sum of these two partial widths will nearly saturate the total width. Also, by assuming that both couplings, $g_{f_1'a_0\pi}$ and $g_{a_1f_0\pi}$, are via the non-strange component in the wave functions of $f_1'$ and $a_1$, they can be expressed by the bare $g_{ASP}$ coupling. 

At this moment the experimental data do not allow a precise determination of these parameters. Our strategy for the parameter determinations is as follows: Since the $K\bar{K}\pi$ channel is dominant in the $f_1'$ decay we assume that this channel accounts for a partial width of 50 MeV such that we can describe the available data best. This determines $g_{f_1'K^*\bar{K}}=2.17$. In the SU(3) flavor symmetry the coupling $g_{f_1K^*\bar{K}}=1.24$ and $g_{a_1K^*\bar{K}}=2.04$ can then be determined. Notice that the couplings $g_{f_0K^+K^-}$ and $g_{a_0K\bar{K}}=2.24$ are relatively well established quantities, and $g_{K^*K\pi}=3.2$ can be determined by experimental data for $K^*\to K\pi$. In Table~\ref{coupling-constants} the coupling constants which appear in the $C=+1$ axial vector meson decays are listed.

\begin{table}
\centering
\caption{Coupling constants adopted in the $1^{++}$ sector. }\label{coupling-constants}
\begin{tabular}{|c|c|}
  \hline\hline
  Coupling constant & Value \\
  \hline
  $g_{f_0K^+K^-}$ & $5.92\pm 0.13$ GeV~\cite{Aloisio:2002bt} \\
  \hline
  $g_{f_0\pi^+\pi^-}$ & $2.96\pm 0.12$ GeV~\cite{Aloisio:2002bt} \\
  \hline
  $g_{a_0K^+K^-}$ & $2.24\pm 0.11$ GeV ~\cite{Aloisio:2002bsa}\\
  \hline
  $g_{a_0\eta\pi}$ & $3.02\pm 0.35$ GeV ~\cite{Aloisio:2002bsa}\\
  \hline
  $g_{AVP}$ & $2.04$ GeV\\
  \hline
  $g_{f_1a_0\pi}$ & $2.93e^{2.26\pi i}$\\
  \hline
  $g_{f_1'a_0\pi}$ & $0.282e^{2.26\pi i}$\\
  \hline
  $g_{f_1K^*\bar{K}}$ & $1.24$ GeV\\
  \hline
  $g_{f_1'K^*\bar{K}}$ & $2.17$ GeV \\
  \hline
  $g_{a_1K^*\bar{K}}$ & $2.04$ GeV \\
  \hline
  $g_{a_1f_0\pi}$ & $2.93e^{2.26\pi i}$ \\
  \hline
\end{tabular}
\end{table}

\subsubsection{Partial decay widths}

In Tables~\ref{partialwidthf1} and \ref{partialwidtha1} we list the calculated partial decay widths for $f_1/f_1'$ and $a_1$, respectively, and compare them with the available data. For the decays of $f_1/f_1'$, it shows that the partial decay widths for both states are consistent with the experimental measurements~\cite{Patrignani:2016xqp}. Some specific points can be learned: 

\begin{itemize}
\item The decay of $f_1\to \eta\pi\pi$ is a dominant channel due to the large tree diagram contribution, i.e. $f_1\to a_0\pi\to \eta\pi\pi$. In contrast, the triangle loop contribution is larger than the tree process in $f_1'\to \eta\pi\pi$. The reason is because of the different couplings in these two channels, i.e. $g_{f_1a_0\pi}>>g_{f_1'a_0\pi}$ and $g_{f_1K^*\bar{K}}<<g_{f_1'K^*\bar{K}}$ (see Table~\ref{coupling-constants}). 

\item The decay of $f_1\to K\bar{K}\pi$ is suppressed by the phase space. In contrast, this is the dominant channel for $f_1'$. The main contribution is from the tree diagram transition of $f_1'\to K^*\bar{K}+c.c.\to K\bar{K}\pi$. Note that the mass of $f_1'$ is located within the kinematic of the TS condition. As a consequence, the contributions from the triangle diagram are larger than the tree process in the transitions involving $a_0\pi$. This also enhances the isospin-breaking effects in $f_1'\to 3\pi$ than the tree-level $a_0-f_0$ mixing contribution. 

\item The isospin-breaking decays, in $f_1$ and $f_1'\to 3\pi$, can be illustrated by the following ratios, respectively, 
\begin{eqnarray}\label{isospin-breaking-rate-f1}
R_{f_1\to 3\pi} &\equiv &\frac{B.R.(f_1\to f_0\pi\to\pi^+\pi^-\pi^0)}{B.R.(f_1\to a_0\pi\to\eta\pi^0\pi^0)}=\frac{3\times 0.01}{9.7}=0.31\% \ ,\nonumber\\
R_{f_1'\to 3\pi} &\equiv &\frac{B.R.(f_1'\to f_0\pi\to\pi^+\pi^-\pi^0)}{B.R.(f_1'\to a_0\pi\to\eta\pi^0\pi^0)}=\frac{3\times 0.16}{10}=4.8\%.
\end{eqnarray}
It shows that the $a_0$-$f_0$ mixing and contributions from the triangle diagram are compatible. Due to the low mass of $f_1$, the condition for the TS mechanism has not been fulfilled. Thus, the isospin-breaking effects in $f_1\to 3\pi$ are  rather small. In contrast, the isospin-breaking effects in $f_1'\to 3\pi$ are strongly enhanced by the TS mechanism. One notices that the exclusive partial width of $f_1'\to 3\pi$ via the tree-level $a_0$-$f_0$ mixing is much smaller than that of $f_1\to 3\pi$. The reason is that the coupling $g_{f_1'a_0\pi}$ is much smaller than $g_{f_1a_0\pi}$ as shown in Tab.~\ref{coupling-constants}. It is interesting to compare the above results with the anomalously large isospin violations in $\eta(1405/1475)\to 3\pi$ ($17.9\%$~\cite{BESIII:2012aa}).

\item In Tab.~\ref{partialwidtha1} the partial decay widths of $a_1^-\to \rho\pi\to 3\pi$ and  $a_1^-\to f_0\pi^-$ are predicted. The $\rho\pi$ channel contributes dominantly and the $f_0\pi$ channel is very small. This is consistent with the experimental measurement, and may reflect the molecular nature of $f_0$.

\end{itemize}

\begin{figure}
  \centering
  \includegraphics[width=3.0in]{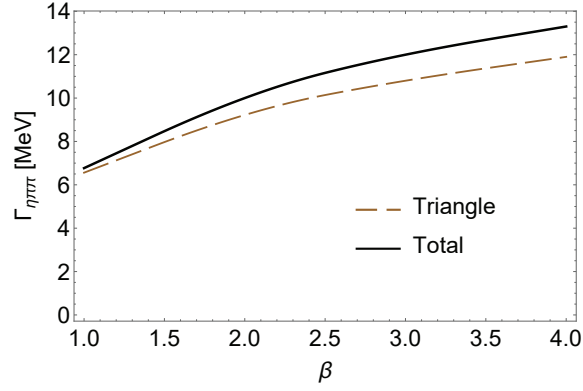}
  \caption{Sensitivity of $\Gamma(f_1'\to a_0\pi\to\eta\pi\pi)$ to the cut-off parameter $\beta$.}\label{cutdepend}
\end{figure}

We also estimate the errors with the partial decay widths which are from two sources. One is the uncertainties with the coupling constants, and the other is due to the cut-off dependence of the loop integrals. For instance, in $f_1'\to K\bar{K}\pi$  we consider the errors with $g_{K^*K\pi}$, $g_{a_0K\bar{K}}$ and $g_{a_0\eta\pi}$ which are extracted from the experimental data (Table~\ref{coupling-constants}) for the first type of the error source. For the second source we find that the results are insensitive to the cut-off energy.  In Fig.~\ref{cutdepend} we plot the cut-off dependence of the partial decay width for $f_1'\to a_0\pi\to\eta\pi\pi$ as a demonstration. It shows that the partial decay width varies rather slowly in terms of the cut-off parameter.

\begin{table}
  \centering
  \caption{ Calculated partial widths of $f_1$ and $f_1'$, in unit of MeV. The corresponding experiment values are shown in round brackets.}\label{partialwidthf1}
  \begin{tabular}{|l|c|c|c|}
    \hline\hline
      & $f_1$ & $f_1'$\\
    \hline
    $a_0\pi\to\eta\pi\pi$ (Fig.~\ref{diagramf1toetapipi} (a): tree) & $7.5\pm1.7$ & $(1.7\pm0.4)\times 10^{-1}$\\
    \hline
    $a_0\pi\to\eta\pi\pi$ (Fig.~\ref{diagramf1toetapipi} (b): tri.) & $(2.0\pm0.6)\times 10^{-1}$ & $9.2\pm3.0$\\
    \hline
    Partial width of $f_1/f_1'\to \eta\pi\pi$ & $9.7\pm2.4$ ($8.6\pm1.3$~\cite{Patrignani:2016xqp}) & $10\pm3.5$ \\
    \hline\hline
    $a_0\pi\to K\bar{K}\pi$ (Fig.~\ref{diagramf1tokkpi} (a): tree) & $(3.7\pm0.4)\times 10^{-1}$ & $(1.7\pm0.2)\times 10^{-2}$\\
    \hline
    $a_0\pi\to K\bar{K}\pi$ (Fig.~\ref{diagramf1tokkpi} (c): tri.) & $(2.4\pm0.5)\times 10^{-2}$ & $1.2\pm0.2$\\
    \hline
    $a_0\pi\to K\bar{K}\pi$ (Fig.~\ref{diagramf1tokkpi} (a)+(c) ) & $(5.0\pm0.5)\times 10^{-1}$ & $1.0\pm0.1$ \\
    \hline
    $K^*\bar{K}\to K\bar{K}\pi$ (Fig.~\ref{diagramf1tokkpi} (b): tree) & $(3.0\pm0.1)\times 10^{-1}$ (not seen~\cite{Patrignani:2016xqp}) & $49\pm1$ \\
    \hline
    Partial width of $f_1/f_1'\to K\bar{K}\pi$ & $1.47\pm0.20$ ($1.95^{+0.37}_{-0.39}$~\cite{Barberis:1998by}, $2.1\pm0.2$~\cite{Patrignani:2016xqp}) & $52\pm3$ \\
    \hline\hline
    $f_0\pi\to 3\pi$ (Fig.~\ref{diagramf13pi} (a): mixing) & $(3.6\pm1.2)\times 10^{-3}$ & $(9.0\pm2.9)\times 10^{-5}$\\
    \hline
    $f_0\pi\to 3\pi$ (Fig.~\ref{diagramf13pi} (b): tri.) & $(1.0\pm 0.1)\times10^{-3}$ & $(1.1\pm0.1)\times 10^{-1}$\\
    \hline
    $f_0\pi\to 3\pi$ (Fig.~\ref{diagramf13pi} (c): mixing via tri.) & $(1.4\pm0.4)\times 10^{-4}$ & $(8.0\pm2.6)\times 10^{-3}$\\  
    \hline
    Partial width of $f_1/f_1'\to f_0\pi\to 3\pi$ & $(1.0\pm0.3)\times 10^{-2}$ & $(1.6\pm 0.3)\times 10^{-1}$\\
    \hline\hline
  \end{tabular}
\end{table}

\begin{table}
  \centering
  \caption{Predicted partial decay widths of $a_1$ in comparison with experimental data from CLEO Collaboration~\cite{Asner:1999kj}. }\label{partialwidtha1}
  \begin{tabular}{|l|c|}
    \hline\hline
    Channel & Width\\
    \hline
    $a_1^-\to \rho\pi\to 3\pi$ & $274$ MeV ($221\pm 5^{+17}_{15}$ MeV)~\cite{Asner:1999kj}\\
    \hline
    $a_1^-\to f_0\pi^-$ (Fig.~\ref{diagram-a1-3pi}(c): tri.) & $0.033$ MeV\\
    \hline\hline
    \end{tabular}
\end{table}

\subsubsection{Invariant mass spectra}

To proceed, we discuss the results for the invariant mass spectra in different transition processes.

{(i) $J/\psi\to \gamma f_1/f_1' \to\gamma \eta\pi\pi$}

\begin{figure}
  \centering
  \includegraphics[width=3.5in]{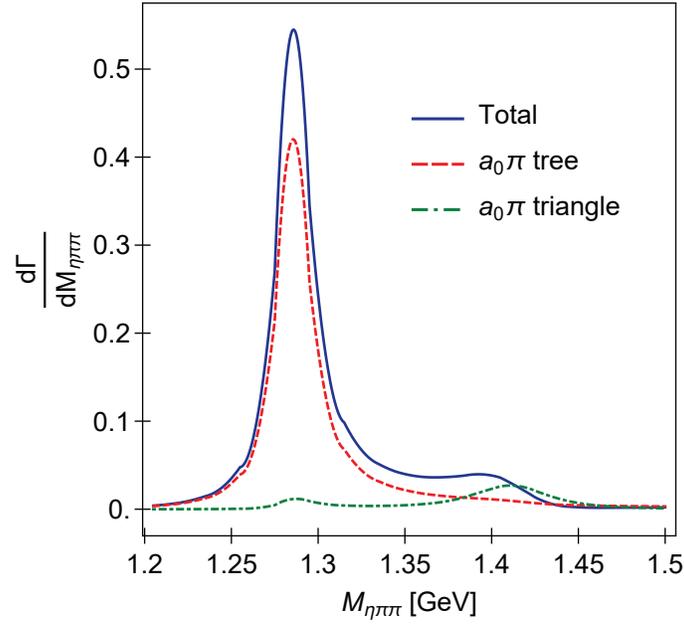}
  \caption{The $\eta\pi\pi$ invariant mass spectrum in $J/\psi\to\gamma f_1/f_1'\to\gamma \eta\pi\pi$. The blue solid, red dashed and green dot-dashed lines are contributions from the full, tree level and triangle amplitudes, respectively.}\label{spectra-f1-etapipi}
\end{figure}

In Fig.~\ref{spectra-f1-etapipi} the $\eta\pi\pi$ invariant mass spectrum (solid line) in $J/\psi\to \gamma f_1/f_1' \to\gamma \eta\pi\pi$ is plotted. As a comparison, the exclusive contributions from the tree-level process $f_1/f_1'\to a_0\pi$ (dashed line) and the triangle loop transition (dot-dashed line) are also presented. It is interesting to see that the signals for $f_1'$ are rather small. This is mainly due to the relatively small production coupling for $J/\psi\to \gamma f_1'$.  One notices that with $\alpha_f=84.48^\circ$ the production coupling ratio in Eq.~(\ref{prod-ratio}) gives ${g_f'}/{g_f}={(\sqrt{2}\cos{\alpha_f}-\sin{\alpha_f})}/{(\sqrt{2}\sin{\alpha_f}+\cos{\alpha_f})}\simeq -0.57$ which suggests that the production of $f_1$ will be relatively enhanced. However, if we only compare the partial decay widths between $f_1$ and $f_1'\to \eta\pi\pi$, they are actually comparable with each other as shown in Table~\ref{partialwidthf1}.

{(ii) $J/\psi\to \gamma f_1/f_1' \to\gamma K\bar{K}\pi$}

\begin{figure}
  \centering
  \includegraphics[width=3.5in]{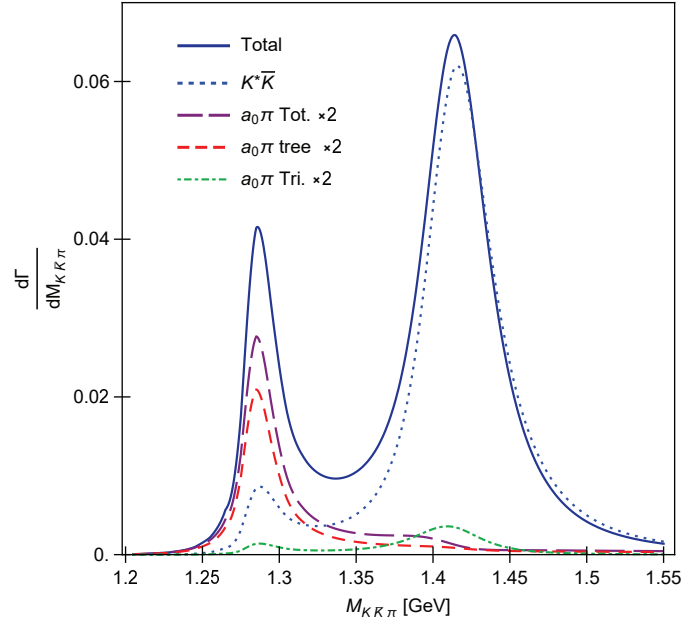}
  \caption{$K\bar{K}\pi$ spectrum of $J/\psi\to\gamma f_1/f_1'\to\gamma K\bar{K}\pi$. The full spectrum is represented by the blue solid line. The blue dotted line stands for the tree-level $K^*\bar{K}$ contribution. The red short dashed and green dot-dashed lines are contributions from the tree-level and  triangle amplitudes of the $a_0\pi$ channel, respectively, and the purple long dashed is the sum of both. For a better display, the latter three contributions are multiplied by a factor of 2. }\label{spectra-f1-KKbarpi}
\end{figure}

In Fig.~\ref{spectra-f1-KKbarpi} the invariant mass spectrum of $K\bar{K}\pi$ in $J/\psi\to\gamma f_1/f_1'\to\gamma K\bar{K}\pi$ is plotted where the signals for both $f_1$ and $f_1'$ are clear. Recall that the production of $f_1'$ in the $J/\psi$ radiative decays will be relatively suppressed. The strong signals of $f_1'$ are driven by its strong coupling to $K^*\bar{K}$ which leads to the dominant contributions from the tree diagram of Fig.~\ref{diagramf1tokkpi} (b), i.e. $f_1'\to K^*\bar{K}\to K\bar{K}\pi$. The exclusive contributions are shown by the dotted line in Fig.~\ref{spectra-f1-KKbarpi}. In contrast, the intermediate $a_0\pi$ contributions are negligible.

\begin{figure}
  \centering
  \includegraphics[width=3.2in]{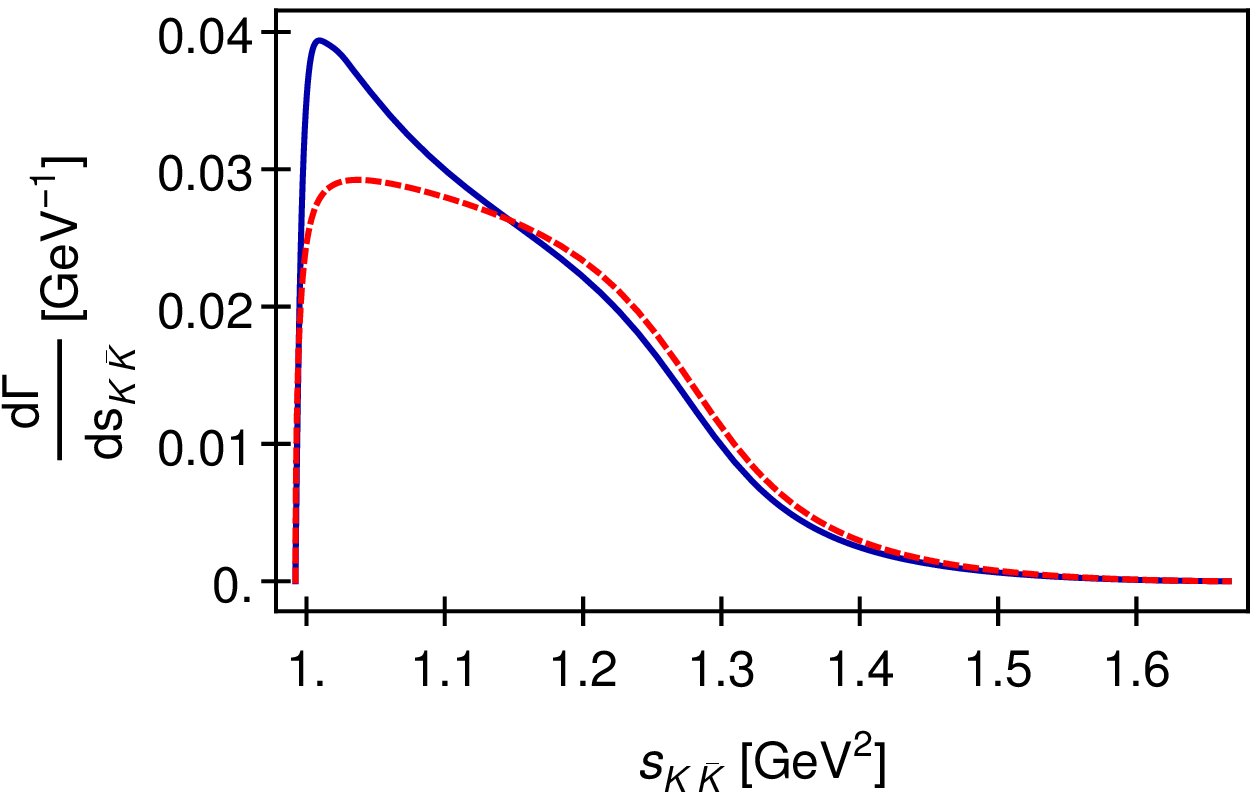}
  \includegraphics[width=3.2in]{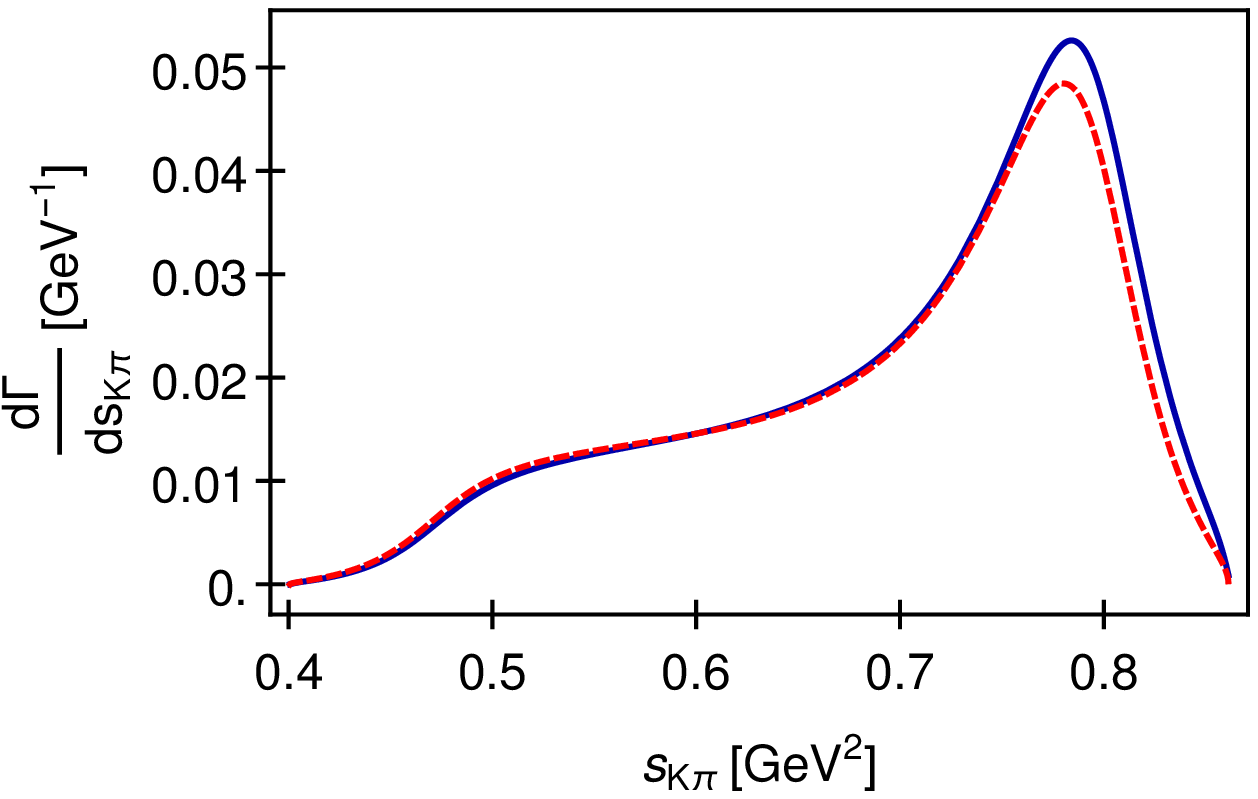}
  \caption{Invariant mass spectra of $K\bar{K}$ (left) and $K\pi$ (right) in $J/\psi\to\gamma f_1'\to\gamma K\bar{K}\pi$. The blue solid lines denote the full calculation results, and the red dashed lines denote the contribution from the tree-level $K^*\bar{K}$ channel.}\label{kkkpispctra}
\end{figure}

In Fig.~\ref{kkkpispctra} the invariant mass spectra of $K\bar{K}$ (left panel) and $K\pi$ (right panel) in $J/\psi\to \gamma f_1' \to\gamma K\bar{K}\pi$ are illustrated by the solid lines. The dashed lines denote the exclusive contributions from the tree diagram of $f_1'\to K^*\bar{K}\to K\bar{K}\pi$. The difference between the solid and dashed lines indicates the effects from the intermediate $a_0\pi$ contributions. In particular, for the $f_1'$ decays into $K\bar{K}\pi$ the TS contribution is larger than the tree-level $a_0\pi$ contribution. Therefore, the difference between the solid and dashed lines is largely due to the TS  mechanism. Meanwhile, the strong threshold enhancement at the $K\bar{K}$ threshold is due to the $a_0(980)$ pole structure. In contrast, in Fig.~\ref{kkkpispctra} (b) the signal for $K^*$ is evident.

{(iii) $J/\psi\to \gamma f_1/f_1' \to\gamma 3\pi$}

In this process the decays of $f_1/f_1'$ into $3\pi$ violate isospin symmetry. For the processes in Fig.~\ref{diagramf13pi} the isospin-violating mechanisms actually imply an abnormal lineshape with the $\pi\pi$ invariant mass spectrum due to the cancellations outside the kinematic regions between the thresholds of the charged and neutral $K\bar{K}$ pairs.

\begin{figure}
  \centering
  \includegraphics[width=3.2in]{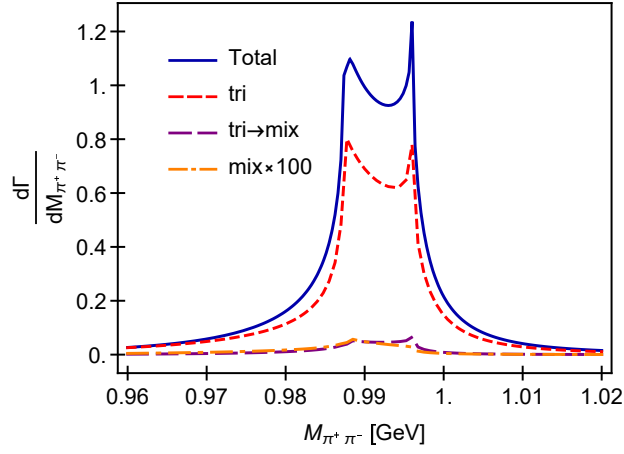}
  \caption{Invariant mass spectrum of $\pi^+\pi^-$ in $f_1'\to \pi^+\pi^-\pi^0$. The dot-dashed, short dashed and long dashed lines correspond to the contributions from Fig.~\ref{diagramf13pi} (a), (b) and (c), respectively.}\label{pipispectrum}
\end{figure}

The $\pi^+\pi^-$ invariant mass spectrum is shown in Fig.~\ref{pipispectrum}, where we can see the cancellation between charged and neutral loop amplitudes produces a narrow $f_0(980)$ peak with a width of $\sim(2m_{K^0}-2m_{K^+})\simeq 10$ MeV. In general, this is the signature for the $a_0$-$f_0$ mixing mechanism for the isospin breaking effects. This narrow structure also becomes the signature for the TS mechanism if significantly large isospin breaking effects are observed within the TS kinematic region~\cite{Wu:2011yx,Wu:2012pg,Du:2019idk}.

Some special features arise from the process of $J/\psi\to \gamma f_1/f_1' \to\gamma 3\pi$. Since the tree diagram (Fig.~\ref{diagramf13pi} (a)) is relatively suppressed by the $f_1-f_1'$ mixing the isospin violation of $f_1'$ via the tree-level $a_0-f_0$ mixing is also suppressed. As shown in Table~\ref{partialwidthf1} the TS mechanism has the largest contribution to the isospin breaking effects. In addition, the TS mechanism can enhance the production of $a_0$ which will further enhance the $a_0-f_0$ mixing contributions. This scenario is similar to the case of the isospin violation of $\eta(1405/1475)\to 3\pi$~\cite{Du:2019idk}, but have been overlooked in other studies~\cite{Wu:2011yx,Wu:2012pg,Aceti:2012dj,Achasov:2015uua}. One can see that although this amplitude is small, its interferences cannot be neglected (see Table~\ref{partialwidthf1}). Combining together the isospin-breaking contributions from those three processes in Fig.~\ref{diagramf13pi}, we see that the isospin breaking effects are about $4.8\%$ (see Eq.~(\ref{isospin-breaking-rate-f1})). This value is significantly larger than the usual ones of about $1\sim 2\%$ from the pure $a_0-f_0$ mixing, and has indicated a signature of the TS mechanism.  

We will also see later that the TS mechanism will put a constraint on the $a_1(1420)\to 3\pi$. With the same initial energy around 1.4 GeV, the charged and neutral loop amplitudes will constructively add to each other.

{(iv) $\chi_{c1}\to a_1\pi\to 4\pi$}

\begin{figure}
  \centering
  \includegraphics[width=3.2in]{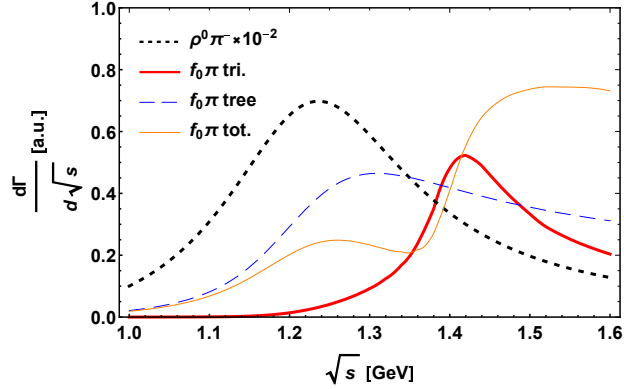}
  \caption{The invariant mass spectrum of $\pi^+\pi^-\pi^-$ in $\chi_{c1}\to a_1(1260)\pi\to 4\pi$. The black dotted, red thick solid, blue dashed and orange thin solid lines represent the contributions from the tree diagram of the $\rho^0\pi^-$ channel, triangle diagram of the $f_0\pi$ channel, tree diagram of the $f_0\pi$ channel, and a coherent sum of the $f_0\pi$ channels, respectively. The blue dashed and orange thin solid line are calculated assuming $\alpha_{f0}=\pm\frac{\pi}{2}$. Note that the $\rho\pi$ contributions are scaled down by a factor of $10^{-2}$, which means that the contributions from the intermediate $f_0\pi$ channels are small.}\label{a1tof0pispctra}
\end{figure}

As shown in subsection~\ref{sect:a1} the decay of $\chi_{c1}\to 4\pi$ gives access to the production of $a_1(1260)$ in its decays into $3\pi$. Note that $a_1(1260)$ is a broad state, i.e. $\Gamma_{a_1}=250\sim 600$ MeV~\cite{Patrignani:2016xqp}, it is necessary to investigate the $3\pi$ invariant mass spectrum in a relatively broad mass region. This will bring the kinematics into the physical region of the TS condition. Therefore, a systematic study of the $a_1(1260)$ production in a broad mass region via the $3\pi$ invariant mass spectrum will provide a test of the nature of $a_1(1420)$ which could be an enhancement produced by the TS mechanism.

The corresponding processes are illustrated in Fig.~\ref{diagram-a1-3pi}. In Fig.~\ref{a1tof0pispctra} the $3\pi$ invariant mass spectrum is plotted. One can see that the dominant decay of $a_1(1260)$ is the $\rho\pi$ channel, which accounts for about $60\%$ of the total width~\cite{Patrignani:2016xqp}. This helps determine the coupling of $a_1\to\rho\pi$. As shown by the dashed line, the intensity of $f_0\pi^-$ channel relative to the $\rho^0\pi^-$ channel ranges from $(0.4\sim 1)\%$. This is consistent with the COMPASS observation~\cite{Adolph:2015pws}. Given the broad width of $a_1(1260)$ and its strong coupling to the nearby $K^*\bar{K}+c.c.$ threshold, the TS mechanism can produce a significant enhancement at about 1.4 GeV (see the solid line). This structure, noted as $a_1(1420)$, can be regarded as a natural consequence of the TS mechanism and is not necessarily interpreted as a genuine resonance state. The orange thin solid line in Fig.~\ref{a1tof0pispctra} shows the combined contributions from $a_1(1260)\to f_0\pi$. The lineshape indicates the interference between the amplitudes of the tree-level and TS transitions, and is consistent with the partial wave analysis in experiment~\cite{Adolph:2015pws}. The observation of $a_1(1420)$ hence can be regarded as a signature for the TS mechanism which drives the rich phenomena observed near the $K^*\bar{K}+c.c.$ threshold in various processes.

\subsection{Numerical results for the $J^{PC}=1^{+-}$ states}

\subsubsection{Parameters}

Parameters in the calculations of the $C=-1$ axial vectors include its production coupling in $J/\psi$ decays into a pseudoscalar and $C=-1$ axial vector, i.e. $g_{\psi BP}$. As discussed in Subsection~\ref{prod-negative-c}, this quantity can be determined by the experimental data for $J/\psi\to b_1^{\pm}\pi^{\mp}$~\cite{Patrignani:2016xqp}. The experimental  data for $h_1\to\rho\pi$ and $b_1^+\to \omega\pi^+$ can connect all the couplings among the SU(3) multiplets together as discussed in subsections~\ref{prod-negative-c} and \ref{sub:b1toKKpi}. In Table~\ref{coupling-constants-B} all the vertex couplings are listed. 

\begin{table}
\centering
\caption{Coupling constants for the $1^{+-}$ sector.}\label{coupling-constants-B}
\begin{tabular}{|c|c|}
  \hline\hline
  Coupling constant & Value \\
  \hline
  $g_{BVP}$ & $3.03$ GeV\\
  \hline
  $g_{h_1\rho\pi}$& $4.28$ GeV\\
  \hline
  $g_{h_1'\rho\pi}$& $0.13$ GeV\\
  \hline
  $g_{h_1'K^*\bar{K}}$& $3.09$ GeV\\
  \hline
  $g_{b_1K^*\bar{K}}$& $3.03$ GeV \\
  \hline
  $g_{b_1\omega\pi}$& $4.28$ GeV\\
  \hline
  $g_{\psi BP}$ & $4.35\times 10^{-3}$ GeV\\
  \hline\hline
\end{tabular}
\end{table}

\subsubsection{Invariant mass spectra}

(i) $J/\psi\to b_1\pi \to \phi\pi^+\pi^-$
 
\begin{figure}
  \centering
  \includegraphics[width=3.2in]{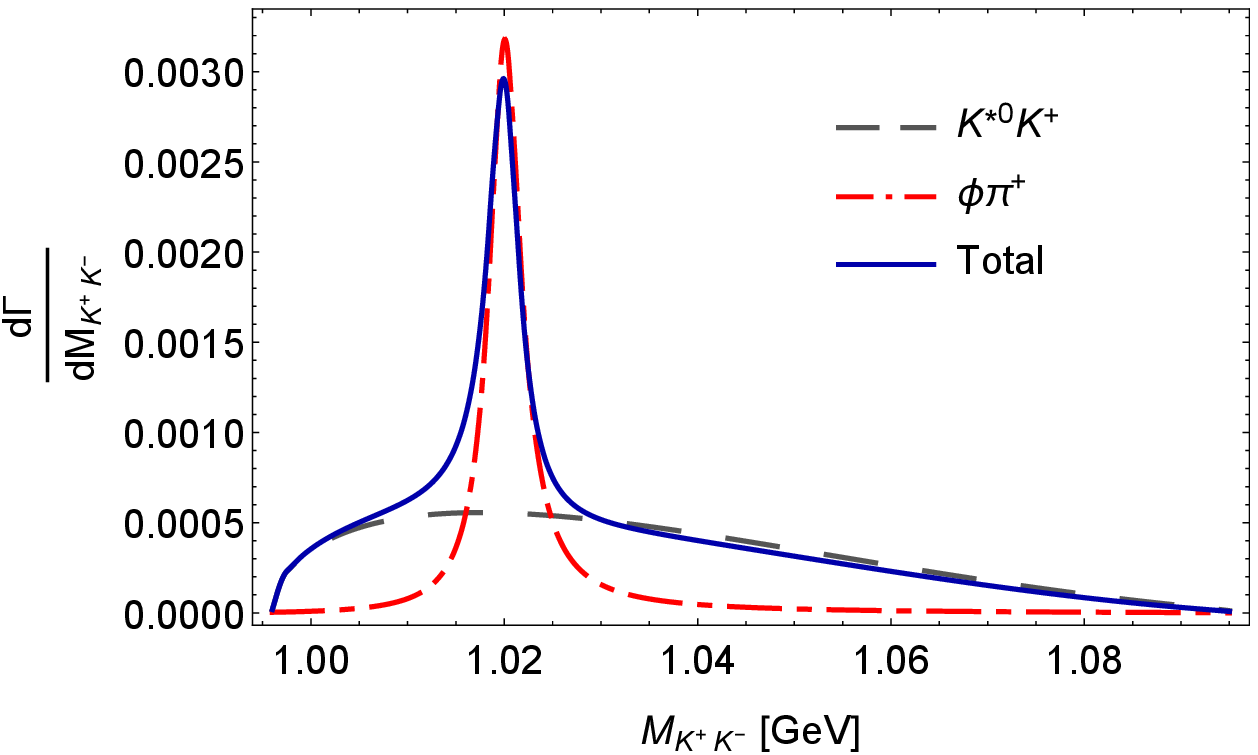}
  \includegraphics[width=3.2in]{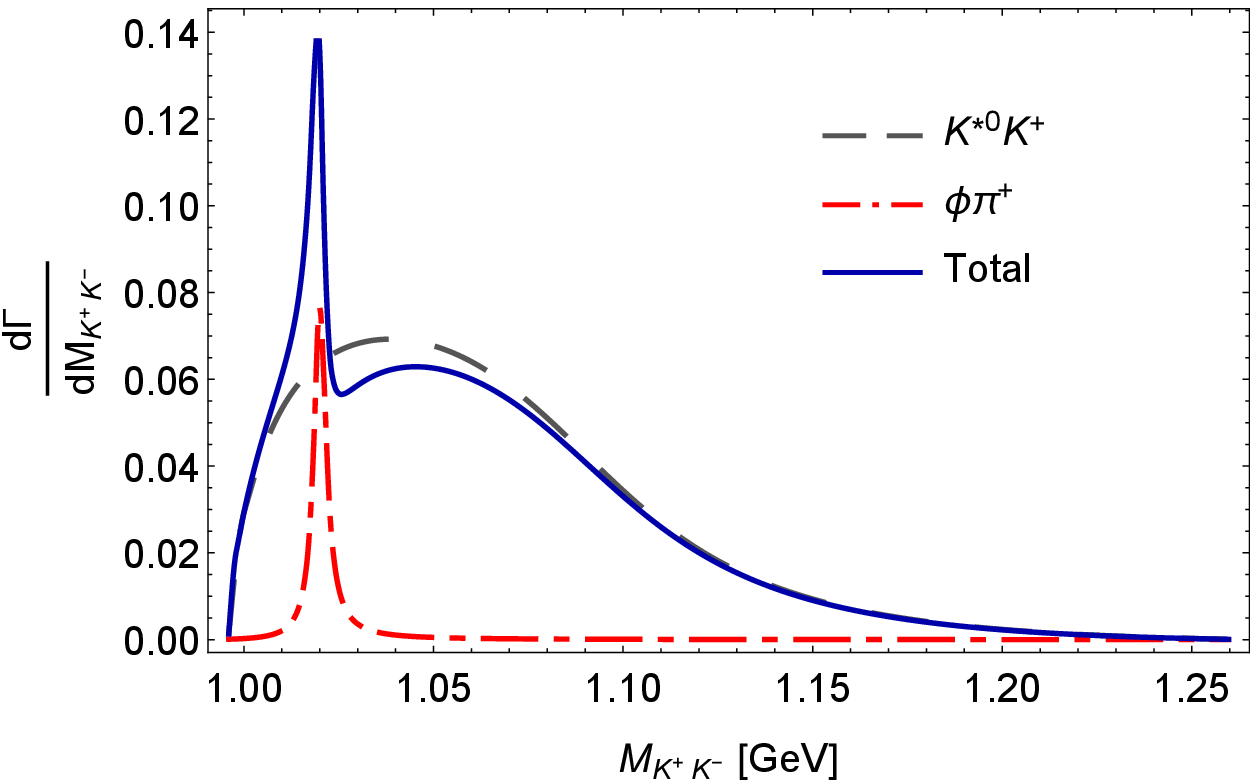}
  \caption{Invariant mass spectrum of $K^+K^-$ in  $J/\psi\to b_1^+\pi^-\to K^+K^-\pi^+\pi^-$ at $\sqrt{s}=m_{b_1}$ (left) and $\sqrt{s}=1.4$ GeV (right). The gray dashed,  red dot-dashed and blue solid lines represent the contributions from the $K^+\bar{K}^{*0}$ , $\phi\pi^+$ channels and the coherent sum, respectively.\label{spectrum_KK_from_b1_phipiandKstrK_KKpi}}
\end{figure}

As discussed earlier the decay of $b_1\to \phi\pi$ is an OZI violation process. The triangle loop actually provides a mechanism to evade the OZI rule.  The OZI-rule violation effects can be indicated by the following branching ratio fraction: 
\begin{eqnarray}
R_{b_1}\equiv\frac{B.R.(b_1^+\to\phi\pi^+\to K\bar{K}\pi^+)}{B.R.(b_1^+\to\omega\pi^+)(S-wave)}=\frac{\Gamma(b_1^+\to\phi\pi^+\to K^0\bar{K}^0\pi^+)+\Gamma(b_1^+\to\phi\pi^+\to K^+K^-\pi^+)}{\Gamma(b_1^+\to\omega\pi^+)}=1.8\times 10^{-4} .
\end{eqnarray}
which is consistent with the experimental upper limit $R_{b_1}<0.004$. Recalling that $b_1$ is broad and the decay of $b_1\to \phi\pi\to K\bar{K}\pi$ will have the tree-level contribution from $b_1\to K^*\bar{K}\to K\bar{K}\pi$ as the background contribution, it should be interesting to examine the $K\bar{K}$ invariant mass spectrum from the on-shell mass to the TS kinematic region in the $J/\psi$ decay. In Fig.~\ref{spectrum_KK_from_b1_phipiandKstrK_KKpi} the $K^+K^-$ spectra at $\sqrt{s}=m_{b_1}=1.235$ GeV (left) and $\sqrt{s}=1.40$ GeV (right) are plotted with the cut-off parameter $\beta=2$. Although the $\phi$ meson production is an OZI-rule violating process the signal stands clearly out of the smooth background. Comparing these two plots in Fig.~\ref{spectrum_KK_from_b1_phipiandKstrK_KKpi} it shows that the lineshape of the $\phi$ meson is distorted at $\sqrt{s}=1.40$ GeV due to the TS mechanism. Moreover, the cross section at $\sqrt{s}=1.40$ GeV is much larger than that at $\sqrt{s}=m_{b_1}=1.235$ GeV. This is a strong indication of the TS mechanism in this channel.

\begin{figure}
  \centering
  \includegraphics[width=3.2in]{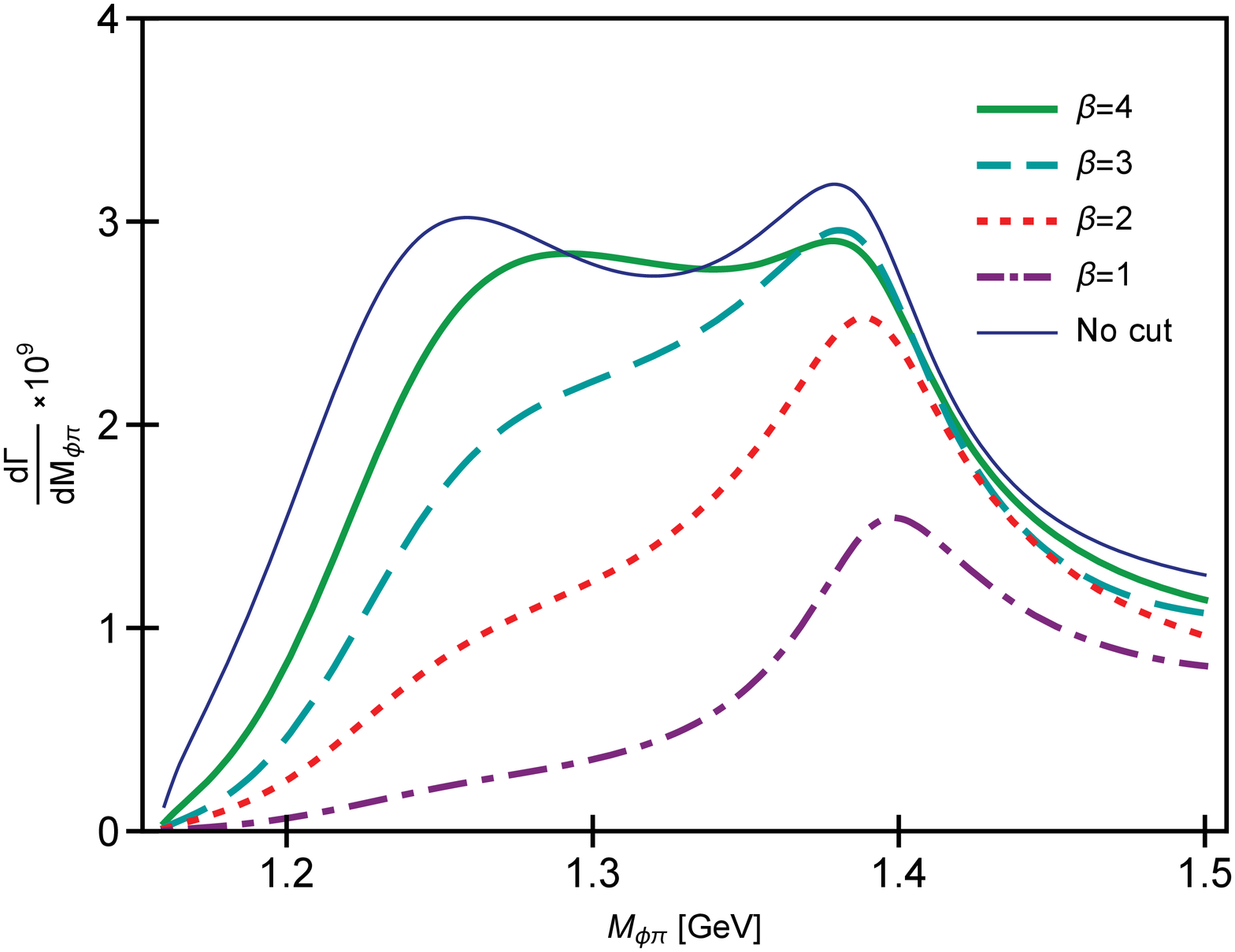}
  \includegraphics[width=3.2in]{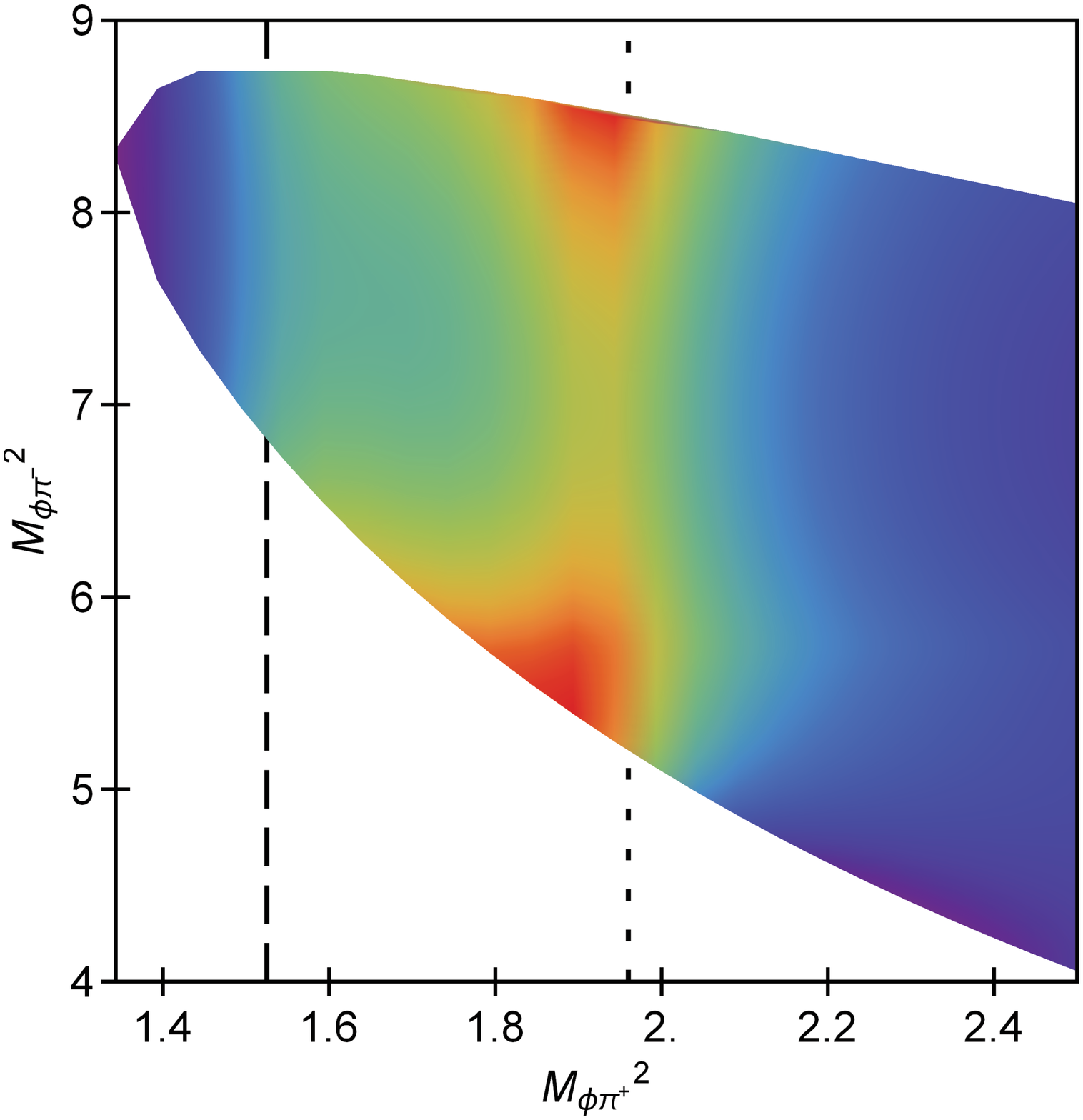}
  \caption{Invariant mass spectrum of $\phi\pi^+$ (left) and Dalitz plot of $J/\psi\to b_1^+\pi^-+c.c.\to\phi\pi^+\pi^-$ (right). In the $\phi\pi$ spectrum (left), the green solid, cyan dashed, red dotted, purple dot-dashed, and blue thin-solid lines stands for calculations with different cut-off parameters, i.e. $\beta=4$, $\beta=3$, $\beta=2$, $\beta=1$, and without the form factor, respectively. The first peak is due to the resonance of $b_1(1235)$  and the second peak is due to the presence of the TS. The Dalitz plot (right) is calculated with $\beta=2$. The mass of $b_1(1235)$ and the location of the TS are marked by the dashed and dotted lines, respectively.}\label{phipipispectrum}
\end{figure}

In Fig.~\ref{phipipispectrum} we present the $\phi\pi^+$ invariant mass spectrum and the Dalitz plot for $\phi\pi^+\pi^-$ which can further disentangle the role played by the TS mechanism. In Fig.~\ref{phipipispectrum} (a), a set of curves which correspond to different cut-off energies ($\beta\Lambda_{QCD}$), are shown. It is interesting to see that although the $b_1$ signal is sensitive to the cut-off energies, the TS peak (close to the normal threshold of $K^*\bar{K}$) appears to be stable. This is because, at the kinematic region away from the on-shell conditions, the loop integrals, which are dominated by the dispersive part, will become more sensitive to the cut-off energies. In contrast, at the TS kinematics the internal states are all on-shell or near threshold. The absorptive part is dominant and the cut-off becomes quite irrelevant. The dispersive part, when moves away from the on-shell kinematics, will always be largely cut off. Therefore, the residue amplitude is insensitive to the cut-off energies. 

With a typical cut-off energies $\beta\Lambda_{QCD}=500$ MeV (i.e. $\beta=2$), we see that the red dotted line in Fig.~\ref{phipipispectrum} (a) has a trivial structure at the mass of $b_1$, but a clear TS enhancement at the $K^*\bar{K}$ threshold. This mechanism can produce signature pattern in the Dolitz plot as shown by Fig.~\ref{phipipispectrum} (b). The presence of the TS mechanism actually makes a unique prediction for the $\phi\pi^+$ invariant mass spectrum in $J/\psi\to b_1 \pi\to \phi\pi^+\pi^-$ that can be investigated in experiment.

(ii) $J/\psi\to h_1'\eta \to\phi\pi\eta\to K\bar{K}\pi\eta$

The decay channel of $J/\psi\to h_1'\eta \to\phi\pi\eta\to K\bar{K}\pi\eta$ can also provide signals for the TS mechanism. Although the mass of $h_1'$ still bares a relatively large uncertainty, the signature of the TS mechanism exhibits some model independent feature that can be identified in experiment. Note that the decay of $h_1'\to\phi\pi$ is an isospin-violating transition. The TS mechanism actually provides a leading source for violating the isospin symmetry~\footnote{The $\rho^0$-$\phi$ mixing is negligibly small. Thus, we do not include its contributions in this calculation.}.  As discussed earlier, the charged and neutral triangle loop amplitudes will cancel out. This process is correlated with the decay of  $J/\psi\to b_1 \pi\to \phi\pi^+\pi^-$ where the charged and neutral loop amplitudes add constructively to each other.

\begin{figure}
  \centering
  \includegraphics[width=3.2in]{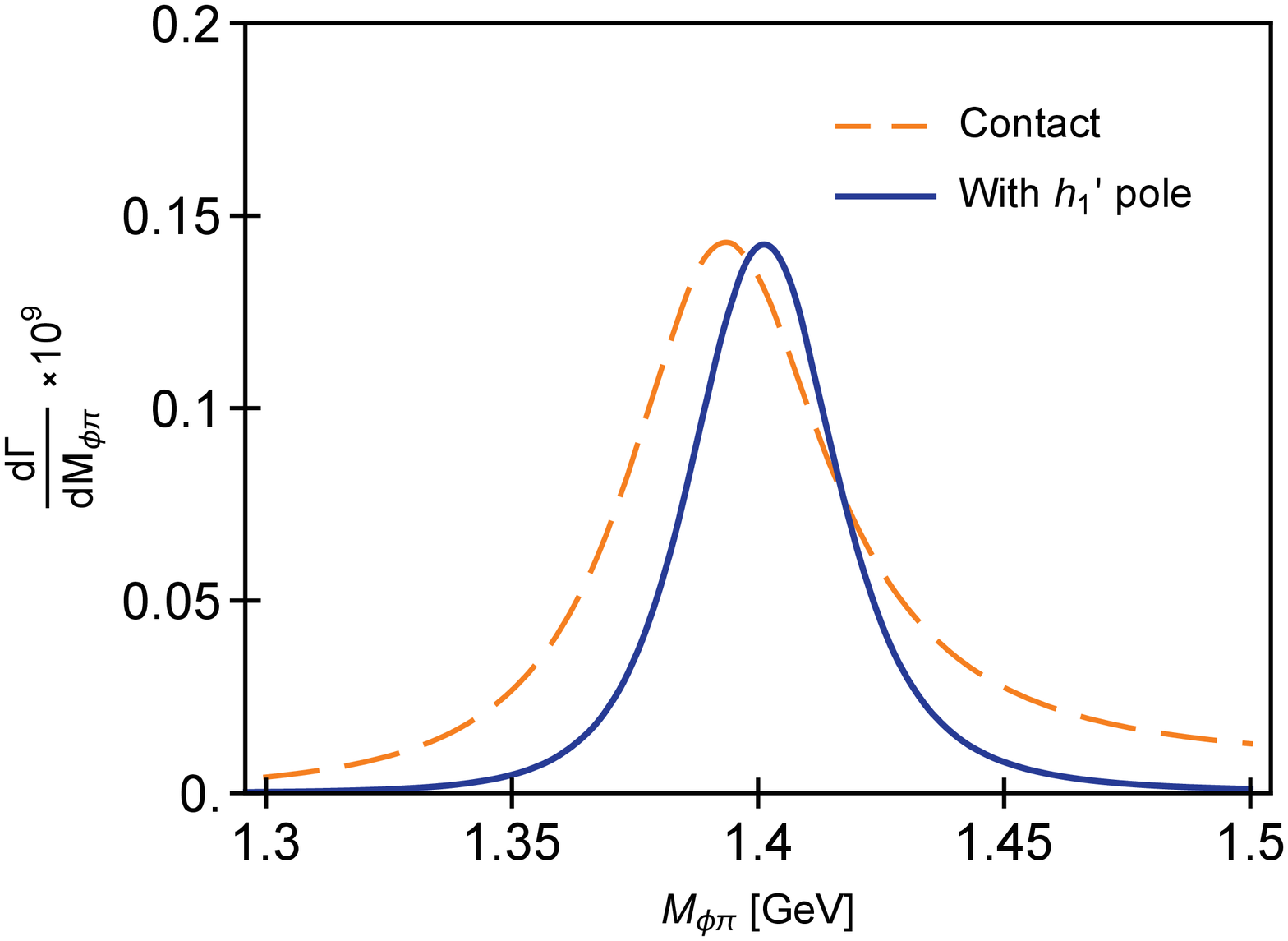}
  \includegraphics[width=3.2in]{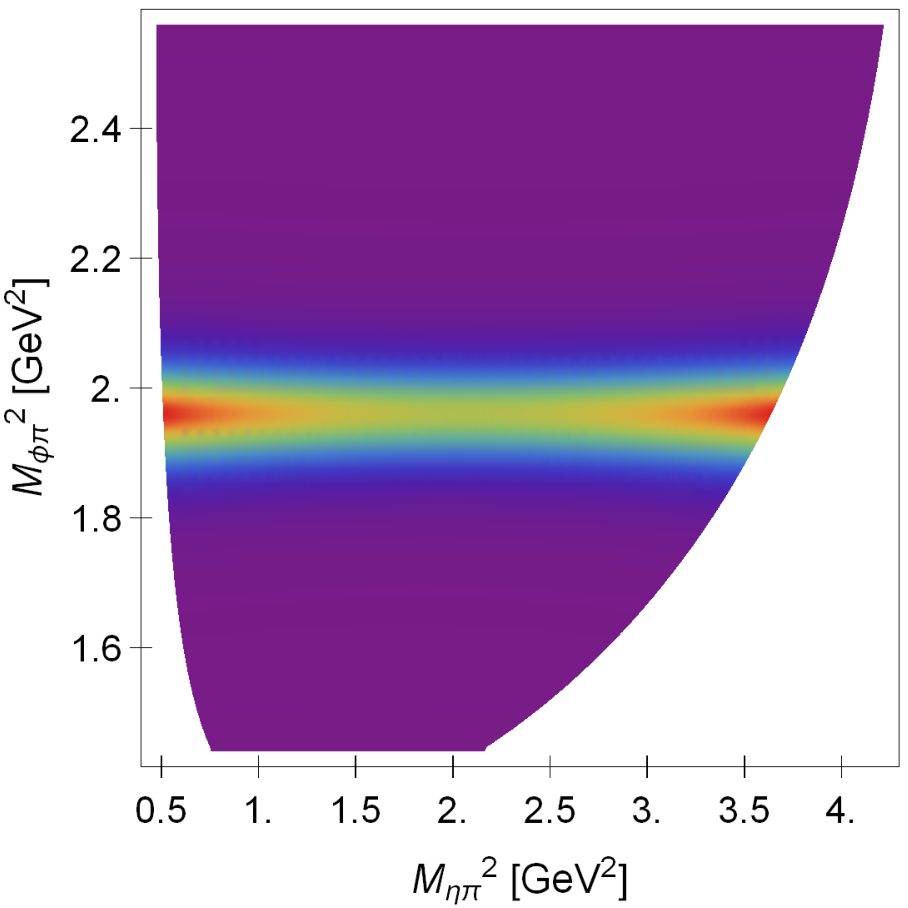}
  \caption{The $\phi\pi$ spectrum of $h_1'$ in $J/\psi\to\eta h_1'\to\eta\phi\pi$ and the Dalitz plot of $J/\psi\to\eta h_1'\to\eta\phi\pi$ (right). The orange dashed line is calculated assuming that there is no $h_1'$ resonance produced in $J/\psi$ decay, and it is normalized to the blue solid one, which is calculated considering the resonance of $h_1'$, with the mass $m_{h_1'}=1.423$ GeV measured by BESIII~\cite{Ablikim:2018ctf}.}\label{etaphipidalitzh1}
\end{figure}

In Fig.~\ref{etaphipidalitzh1} (a) we depict the $\phi\pi^0$ invariant mass spectrum with (blue solid line) and without (orange dashed line) the $h_1'$ propagator. It shows that without the $h_1'$ the cross section will be much smaller and the location of the peak position is determined by the TS kinematics, i.e. between the $K^{*+}K^-$ and $K^{*0}\bar{K}^0$  thresholds. In contrast, the inclusion of the $h_1'$ pole will significantly enhance the peak strength. Furthermore, the combined effects from the pole structure with $m_{h_1'}=1.423$ GeV~\cite{Ablikim:2018ctf} and the TS enhancement will shift the peak position by about 10 MeV towards to the $K^*\bar{K}$ threshold.  It should also be noted that the results are insensitive to the cut-off energies since the dispersive part will cancel out in this isospin-violating process. 

The process $J/\psi\to h_1'\eta\to\phi\pi\eta$ has also been studied in Ref.~\cite{Jing:2019cbw}. Different from our approach, in Ref.~\cite{Jing:2019cbw} no $h_1'$ pole is included and the intermediate $K^*\bar{K}$ is introduced by a contact interaction in the $J/\psi$ decays. As shown by Fig.~\ref{etaphipidalitzh1} (a) without considering the relative magnitude between the two curves it may be difficult to distinguish the two cases by about 10 MeV difference. The Dalitz plot are also shown in Fig.~\ref{etaphipidalitzh1} (b), which is in agreement with the observation of BESIII~\cite{Ablikim:2018pik}. 

\begin{figure}
\centering
\includegraphics[width=3.2in]{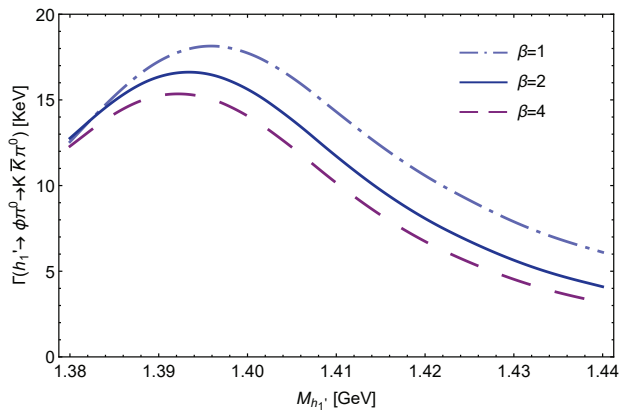}
\caption{The partial width of $h_1'\to\phi\pi\to K\bar{K}\pi^0$ as a function of $M_{h_1'}$. The light blue dot-dashed, blue solid and purple dashed lines represent the calculations with different cut-off energies represented by $\beta$, i.e. $\Lambda_i=m_i+\beta\Lambda_{QCD}$, where $m_i$ is the mass of internal particles of the triangle diagram.\label{width_h1_phipi_KKpi}}
\end{figure}

Taking into account the mass uncertainties with $h_1'$, we also investigate the strength of isospin violations in a range of possible masses. By varying the $h_1'$ mass we plot the partial widths of $h_1'\to\phi\pi\to K\bar{K}\pi^0$ in Fig.~\ref{width_h1_phipi_KKpi} for three different cut-off energies. Due to the enhancement of the TS mechanism, the partial width reaches the maximum near $M_{K\bar{K}\pi}=1.39$ GeV, which is between the $K^{*0}\bar{K}^0$ and $K^{*+}K^-$ thresholds. Also, it shows that the partial width is insensitive to the cut-off energies as shown by the dashed and dot-dashed lines.

It is interesting to compare the branching ratios between $J/\psi\to h_1'\eta\to\phi\pi^0\eta$ and $J/\psi\to b_1^+\pi^-+c.c.\to\phi\pi^+\pi^-$ with the presence of the TS mechanism. Recalling that the former is isospin breaking and the later is OZI evading. With the cut-off parameter $\beta=2$, we find $B.R.(J/\psi\to h_1'\eta\to\phi\pi^0\eta)=6.3\times 10^{-8}$ and $B.R.(J/\psi\to b_1^+\pi^-+c.c.\to\phi\pi^+\pi^-)=1.0\times 10^{-5}$. It shows that the OZI-evading b.r. is much larger than that of the isospin breaking one. This is an interesting indication of the importance of the non-perturbative mechanism via meson loop interactions.

\section{Conclusions}

In this work we present a comprehensive study of the $C=\pm 1$ light axial vector mesons. We first discuss the mixing angles between $f_1$ and $f_1'$ (i.e. $f_1(1285)$ and $f_1(1420)$), and between $h_1$ and $h_1'$ (i.e. $h_1(1170)$ and $h_1(1415)$). These two angles can be related to the mixing between $K_1(1270)$ and $K_1(1400)$ through the Gell-Mann-Okubo mass relation. We then introduce the intermediate $K^*\bar{K}+c.c.$ meson loop transitions in the description of the productions and decays of these axial vector mesons. The presence of the nearby $S$-wave $K^*\bar{K}+c.c.$ to which these axial vector mesons have strong couplings, turns out to be crucial for understanding many puzzling questions related to their productions and decays. This is because that the $K^*\bar{K}+c.c.$ rescatterings by the kaon exchange satisfy the triangle singularity condition. Therefore, the TS mechanism can introduce special interference effects in the exclusive decays of these light axial vector mesons. 

In the $C=+1$ sector the main decay channels of the axial vectors are $K\bar{K}\pi$ and $\eta\pi\pi$ for $f_1$ and $f_1'$, and $3\pi$ for $a_1(1260)$. We show that a combined study of these processes by taking into account of the TS mechanism provides clear evidences for the assignment of $f_1$ and $f_1'$ as the mixing states of the flavor singlet and octet. Although $f_1$ can produce an enhancement around 1.4 GeV due to the TS mechanism, without the inclusion of $f_1'$ as a genuine state and the SU(3) partner of $f_1$, it is impossible to understand its productions and decays in both $K\bar{K}\pi$ and $\eta\pi\pi$ channels. We also show that the TS mechanism accounts for the $a_1(1420)$ enhancement which is originated from the strong $a_1(1260)$ coupling to the $K^*\bar{K}+c.c.$ threshold. The same mechanism also accounts for the relatively large isospin breaking effects in $f_1'\to 3\pi$ except that the charged and neutral triangle loop amplitudes have a destructive interfering phase. The combined analysis of these channels provides a self-consistent check of the underlying dynamics. 

In the $C=-1$ sector the experimental information is still sparse. We choose to investigate two special processes, i.e. $J/\psi\to b_1\pi \to \phi\pi^+\pi^-$ and $J/\psi\to h_1'\eta \to\phi\pi\eta\to K\bar{K}\pi\eta$, in order to highlight the role played by the TS mechanism. In the first process the decay of $b_1\to \phi\pi$ is the OZI-rule suppressed. Thus, the tree-level transition is highly suppressed. The triangle diagram actually provides an OZI-evading mechanism which will produce abnormal lineshapes in both $K\bar{K}$ and $\phi\pi$ invariant mass spectra. In particular, a clear enhancement at about 1.39 GeV is predicted in the $\phi\pi$ invariant mass spectrum due to the TS mechanism. The second process is an isospin-breaking decay for $h_1'\to\phi\pi$. It shows that the TS mechanism provides a leading source of the isospin-breaking effects. It is interesting to notice that whether or not to include the $h_1'$ pole does not cause drastic differences in the lineshapes of the $\phi\pi$ invariant mass spectrum. However, the absolute values of the branching ratios in these two scenarios will be very different. Without the inclusion of the $h_1'$ pole, the branching ratio of $J/\psi\to K^*\bar{K}\eta +c.c.\to \phi\pi\eta$ will be much smaller than that of $J/\psi\to h_1'\eta \to\phi\pi\eta$. These predictions can be searched for in experiment at BESIII.

In brief, the productions and decays of these two sets of axial vector mesons have provided important evidences for the role played by the TS mechanism. This combined analysis has helped clarify some crucial issues concerning their identifications and classifications.

\begin{acknowledgments}
This work is supported, in part, by the National Natural Science Foundation of China (Grant Nos. 11425525 and 11521505),  DFG and NSFC funds to the Sino-German CRC 110 ``Symmetries and the Emergence of Structure in QCD'' (NSFC Grant No. 12070131001, DFG Project-ID 196253076), National Key Basic Research Program of China under Contract No. 2020YFA0406300, and Strategic Priority Research Program of Chinese Academy of Sciences (Grant No. XDB34030302).
\end{acknowledgments}

\section*{Appendix}
\begin{appendix}

We include here a pedagogic deduction of the Gell-Mann-Okubo relation. With the definitions of mixing angles in the article, the masses of $K_{1A}$ and $K_{1B}$ are described by
\begin{eqnarray}
m_{K_{1A}}^2=m_{K_1(1400)}^2\cos^2{\theta_{K_1}}+m_{K_1(1270)}^2\sin^2{\theta_{K_1}}\nonumber\\
m_{K_{1B}}^2=m_{K_1(1400)}^2\sin^2{\theta_{K_1}}+m_{K_1(1270)}^2\cos^2{\theta_{K_1}}.
\end{eqnarray}
Under the bases $\tilde{f}_1$ and $\tilde{f}_8$, the mass matrix is
\begin{eqnarray}
\left(
\begin{array}{cc}
\langle \tilde{f}_8|H^2|\tilde{f}_8\rangle & \langle \tilde{f}_8|H^2|\tilde{f}_1\rangle\\
\langle \tilde{f}_1|H^2|\tilde{f}_8\rangle & \langle \tilde{f}_1|H^2|\tilde{f}_1\rangle
\end{array}\right)=
\left(\begin{array}{cc}
m_{\tilde{f}_8}^2 & m_{\tilde{f}_{18}}^2\\
m_{\tilde{f}_{18}}^2 & m_{\tilde{f}_1}^2
\end{array}\right).
\end{eqnarray}
By diagonalizing the above matrix, one obtains the physical masses of $f$ and $f'$:
\begin{eqnarray}
\left(\begin{array}{cc}
m_{f'}^2 & 0\\
0 & m_{f}^2
\end{array}\right)
=
\left(
  \begin{array}{cc}
     \cos{\theta_f} & -\sin{\theta_f}\\
     \sin{\theta_f} & \cos{\theta_f}
  \end{array}
\right)
\left(\begin{array}{cc}
m_{\tilde{f}_8}^2 & m_{\tilde{f}_{18}}^2\\
m_{\tilde{f}_{18}}^2 & m_{\tilde{f}_1}^2
\end{array}\right)
\left(
  \begin{array}{cc}
     \cos{\theta_f} & \sin{\theta_f}\\
     -\sin{\theta_f} & \cos{\theta_f}
  \end{array}
\right) .
\end{eqnarray}
Then, one has
\begin{eqnarray}
m_{\tilde{f}_1}^2&=&\frac{1}{2}[m_f^2+m_{f'}^2-(m_{f'}^2-m_f^2)\cos{2\theta_f}] \ ,\label{m1squared}\\
m_{\tilde{f}_8}^2&=&\frac{1}{2}[m_f^2+m_{f'}^2+(m_{f'}^2-m_f^2)\cos{2\theta_f}] \ ,\label{m8squared}\\
m_{\tilde{f}_{18}}^2&=&-\frac{1}{2}(m_{f'}^2-m_f^2)\sin{2\theta_f}.\label{m18squared}
\end{eqnarray}
From Eqs.~(\ref{m1squared}),~(\ref{m8squared}) and~(\ref{m18squared}), the following relations can be deduced:
\begin{eqnarray}
m_{\tilde{f}_1}^2+m_{\tilde{f}_8}^2&=&m_f^2+m_{f'}^2 \ ,\\
m_{\tilde{f}_{18}}^4&=&m_{\tilde{f}_1}^2m_{\tilde{f}_8}^2-m_{f'}^2m_f^2 \ (m_{\tilde{f}_{18}}^2<0) \ ,\\
\tan{\theta}&=&\frac{m_{\tilde{f}_{18}}^2}{m_f^2-m_{\tilde{f}_8}^2}=\frac{m_{\tilde{f}_8}^2-m_{f'}^2}{m_{\tilde{f}_{18}}^2}.
\end{eqnarray}
The solution to these equations is Eq.~(\ref{tantheta}).

\end{appendix}

\bibliographystyle{unsrt}

\end{document}